\documentclass[twoside]{article} 
\usepackage{times}
\usepackage{amsmath,amssymb,amsfonts,verbatim,bm,graphicx}
\usepackage{enumerate,epstopdf,lscape,enumitem}
\usepackage[titletoc,title]{appendix}
\usepackage[dvipsnames,usenames]{color}
\usepackage[font=scriptsize]{caption}
\usepackage{caption,subcaption,booktabs}
\usepackage{colonequals}
\usepackage[abbrvbib]{jmlr2e}

\definecolor{dodgerblue}{rgb}{0.12,0.56,1.0}
\definecolor{gold}{rgb}{1.0,0.84,0.0}
\definecolor{green}{rgb}{0,0.65,0.32}
\definecolor{darkorange}{rgb}{1.0,0.55,0.0}
\definecolor{darkblue}{rgb}{0.0,0.0,0.55}
\definecolor{orangered}{rgb}{1.0, 0.27, 0.0}
\definecolor{pink}{rgb}{0.95, 0.55, 0.67}

\setlist{noitemsep}

\def\EE{E}

\newcommand{\ash}{{\sc ash}}
\newcommand{\mysmash}{{\sc smash}}
\newcommand{\mfvb}{{\sc mfvb}}
\newcommand{\Ga}{\alpha}

\newcommand{\Gg}{\gamma}

\newcommand{\Gd}{\delta}

\newcommand{\s}{\sigma}

\begin{document}

\title{\textbf{Flexible Signal Denoising via Flexible Empirical Bayes
    Shrinkage}}

\author{\name Zhengrong Xing
\email zhengrong@statistics.uchicago.edu \\ 
\addr Department of Statistics \\
University of Chicago \\
Chicago, IL 60637, USA 
\AND
Peter Carbonetto 
\email pcarbo@uchicago.edu \\
\addr Research Computing Center and Department of Human Genetics \\
University of Chicago \\
Chicago, IL 60637, USA
\AND 
\name Matthew Stephens
\email mstephens@uchicago.edu \\ 
\addr Department of Statistics and Department of Human Genetics \\
University of Chicago \\
Chicago, IL 60637, USA}
\date{}
\maketitle

\jmlrheading{1}{2020}{1--29}{1/19; Revised 9/20}{00/00}{xing20a}{Xing, Carbonetto and Stephens}

\ShortHeadings{Flexible Denoising via Empirical Bayes Shrinkage}{Xing,
  Carbonetto and Stephens}

\begin{abstract}%
Signal denoising---also known as non-parametric regression---is often
performed through shrinkage estimation in a transformed (e.g.,
wavelet) domain; shrinkage in the transformed domain corresponds to
smoothing in the original domain. A key question in such applications
is how much to shrink, or, equivalently, how much to smooth. Empirical
Bayes shrinkage methods provide an attractive solution to this
problem; they use the data to estimate a distribution of underlying
``effects'', hence automatically select an appropriate amount of
shrinkage. However, most existing implementations of Empirical Bayes
shrinkage are less flexible than they could be---both in their
assumptions on the underlying distribution of effects, and in their
ability to handle heterskedasticity---which limits their signal
denoising applications. Here we address this by taking a particularly
flexible, stable and computationally convenient Empirical Bayes
shrinkage method, and we apply it to several signal denoising
problems. These applications include smoothing of Poisson data and
heteroskedastic Gaussian data. We show through empirical comparisons
that the results are competitive with other methods, including both
simple thresholding rules and purpose-built Empirical Bayes
procedures. Our methods are implemented in the R package {\tt smashr},
``SMoothing by Adaptive SHrinkage in R,'' available at
\url{https://www.github.com/stephenslab/smashr}.
\end{abstract}

{\bf Keywords:} Empirical Bayes, wavelets, non-parametric regression,
mean estimation, variance estimation

\section{Introduction}

Shrinkage and sparsity play a key role in many areas of modern
statistics, including high-dimensional regression
\citep{Tibshirani1996Regression}, covariance or precision matrix
estimation \citep{Bickel2008Covariance}, multiple testing
\citep{Efron2004} and signal denoising \citep{Donoho1994Ideal,
  donoho95}. One attractive way to achieve shrinkage and sparsity is
via Bayesian or Empirical Bayes (EB) methods
\citep[e.g.,][]{Efron2002Empirical, Johnstone2004Needles,
  Johnstone2005Empirical, Clyde2000Flexible,
  Daniels2001Shrinkage}. These methods are attractive because they can
adapt the amount of shrinkage to the available data. Specifically, by
learning from the data the distribution of the underlying ``effects''
that are being estimated, EB methods can appropriately adapt the
amount of shrinkage from data set to data set, and indeed from data
point to data point. For example, in settings where the effects are
sparse, but with a long tail of large effects, optimal accuracy is
achieved by strongly shrinking observations that lie near zero while
minimally shrinking the strongest signals \citep{polson2010shrink}.
This form of shrinkage can be achieved by appropriate EB methods.

One area where Bayesian methods for shrinkage have been found to be
particularly effective is in signal denoising
\citep{abramovich1998wavelet, Clyde2000Flexible,
  Johnstone2005Empirical}. Shrinkage plays a key role in signal
denoising, because signal denoising can be accurately and conveniently
achieved by shrinkage in a transformed (e.g., wavelet) domain
\citep{Donoho1994Ideal}. In empirical comparisons
\citep[e.g.,][]{Antoniadis2001Wavelet, Besbeas2004Comparative},
Bayesian methods often outperform alternatives such as simple
thresholding rules \citep{Coifman1995Translationinvariant,
  Donoho1994Ideal}. However, existing software implementations of
Bayesian and EB methods for this problem are limited; for example, the
{\tt ebayesthresh.wavelet} function in the R package {\tt EbayesThresh} 
\citep{johnstone2005ebayesthresh} only implements methods for the
particular case of estimating Gaussian means with constant variance.

Here we show how EB shrinkage can easily be applied to other signal
denoising problems. The key to this generalization is, in essence, to
use a more flexible EB shrinkage method \citep{stephens2017false}
that---among other appealing features---allows for heteroskedastic
variances. This in turn allows it to tackle signal-denoising problems
with heteroskedastic variances. We provide methods and software
implementations for denoising Gaussian means in the presence of
heteroskedastic variance, denoising Gaussian variances, and denoising
Poisson means. These are all settings that are relatively underserved
by existing implementations. Indeed, we are unaware of any existing EB
implementation for wavelet denoising of either the mean or the
variance in the heteroskedastic Gaussian case.  Consistent with
previous studies \citep{Antoniadis2001Wavelet,Besbeas2004Comparative},
we find that the EB methods are more accurate than commonly used
thresholding rules, and, in the Poisson case, competitive with a
dedicated EB method \citep{Kolaczyk1999Bayesian}. Our methods are
implemented in the R package {\tt smashr}
(``SMoothing by Adaptive SHrinkage in R''), available on GitHub
(\url{https://www.github.com/stephenslab/smashr}).

\section{Background}

\def\bhat{\hat{\beta}}
\def\shat{\hat{s}}
\def\bY{\bm{y}}
\def\bZ{\bm{Z}}
\def\bmu{\bm{\mu}}
\def\bp{\bm{p}}
\def\be{\bm{\epsilon}}
\def\tbY{\tilde\bY}
\def\tbe{\tilde\be}
\def\tbmu{\tilde\bmu}
\def\tmu{\tilde\mu}
\def\tY{\tilde{y}}
\def\s{\sigma}
\def\|{\,|\,}
\def\G{\mathcal G}
\def\bx{\bm{x}}
\def\btheta{\bm{\theta}}
\def\bs{\bm{s}}
\def\bsigma{\bm{\sigma}}
\def\bee{\bm{e}}

Here we briefly review EB shrinkage methods, and show how they can be
applied to a simple signal denoising application--- Gaussian data with
constant variance. The mathematical development mirrors 
\citet{Johnstone2005Empirical}.

\subsection{Empirical Bayes Shrinkage}
\label{sec:EB}

Consider observations $\bx = (x_1, \ldots, x_p)$ of underlying
quantities $\btheta = (\btheta_1, \ldots, \btheta_p)$, with Gaussian
errors having standard deviation $\bs = (s_1, \ldots, s_p)$, for which
we assume, for now, are known; that is,
\begin{equation}
\label{eqn:normalmeans}
\bx \,|\, \btheta \sim N_p(\btheta, \Delta) 
\end{equation}
where $\Delta$ is the diagonal matrix with diagonal entries $s_1^2,
\ldots, s_p^2$. Although it is conceptually straightforward to allow
the $s_j$ to vary, in practice most treatments (and software
implementations) assume them to be constant, $s_j = s$, an issue we
return to later. The goal is to estimate $\btheta$. This is sometimes
called the ``normal means'' problem.

Without any assumptions on $\btheta$, the natural estimate for
$\btheta$ seems to be the maximum likelihood estimate $\bx$.  However,
\citet{james1961estimation} showed that more accurate estimates can be
obtained by using ``shrinkage", which essentially reduces variance at
the cost of introducing some bias.

An attractive way to perform shrinkage in practice is to use EB
methods.  These methods assume that $\btheta$ are independent and
identically distributed from some (unknown) underlying distribution,
$g$, which is further assumed to belong to some specified family of
distributions $\G$. Combining this with \eqref{eqn:normalmeans}
yields:
\begin{align}
\label{eqn:ebshrink}
\bx \,|\, \btheta & \sim N_p(\btheta,\Delta), \\
\label{eqn:ebshrink2}
\theta_1, \ldots, \theta_p & \sim^{\text{\em iid}} g(\cdot), \quad g \in \G.
\end{align}
EB methods estimate $\btheta$ in two steps:
\begin{enumerate}
\item Estimate $g$ by maximum likelihood:
\begin{equation*} 
\hat{g} = \arg \max_{g \in \G} L(g),
\end{equation*} 
where
\begin{equation*}
L(g) \colonequals p(\bx | g) =
  \prod_{j=1}^p \int p(x_j \,|\, \theta_j, s_j) \, g(d\theta_j).
\end{equation*}
\item Estimate each $\theta_j$ using its posterior distribution given
  $\hat{g}$,
\begin{equation}
\label{eqn:post}
p(\theta_j \,|\, \bx, \bs, \hat{g}) \propto
  \hat{g}(\theta_j) \, p(x_j \,|\, \theta_j, s_j).
\end{equation}
\end{enumerate}
For example, we estimate $\theta_j$ using the mean of this
posterior distribution. (One can also use the posterior median, which,
if $\hat{g}$ has a point mass at zero, has a ``thresholding'' property
\citep{Johnstone2005Empirical}. However, we have not found this
necessary to achieve accurate performance in practice.

A key feature of EB methods is that, by estimating $g$ from the data,
they can adapt to each individual data set, essentially learning {\em how
much} to shrink from the available data.

Different EB approaches differ in their assumptions on the family
$\G$, and which assumptions are most appropriate may depend on the
setting. In many settings, including those of interest here, it is
anticipated that $\theta$ may be ``sparse'', with many entries at or
near zero. This can be captured by restricting $\G$ to
``sparsity-inducing'' distributions that are unimodal at zero. For
example, the {\tt EbayesThresh} package
\citep{johnstone2005ebayesthresh} implements two options: (i) $g$
is a mixture of a point mass at zero and a Laplace (or double
exponential) distribution; (ii) $g$ is a mixture of a point mass at
zero and a Cauchy distribution. Another common assumption is that $g$
is a mixture of a point mass at zero and a zero-mean Gaussian
distribution, sometimes referred to as a ``spike and slab'' prior
\citep{Clyde2000Flexible}.

Here we use the flexible ``adaptive shrinkage'' (\ash{}) EB methods
introduced in \citet{stephens2017false}.  These methods allow for more
flexible distributional families $\G$ while maintaining
sparsity-inducing behaviour, and allow the standard deviations $s_j$
to vary. They are also computationally stable and efficient. At its
most flexible, \ash{} assumes $\G$ to be the family of all unimodal
distributions (with their modes set to zero in settings where sparsity
is desired). Here we adopt a slightly more restrictive family, in
which $\G$ is the family of zero-centered scale mixtures of
normals. In practice, this is achieved by using finite mixtures with a
potentially large number of components; that is,
\begin{equation}
  \label{eqn:gnorm}
g(\cdot) = \sum_{k=0}^K \pi_k N(\,\cdot\,;0,\omega_k^2),
\end{equation}
where the mixture weights $\pi_0, \ldots, \pi_K$ are non-negative and
sum to 1, and $N(\,\cdot\,; \mu, \sigma^2)$ denotes the density of the
normal distribution with mean $\mu$ and variance $\sigma^2$.

A key idea, which substantially simplifes inference, is to take
$\omega_0, \ldots, \omega_K$ to be a fixed grid of values ranging from
very small (e.g., $\omega_0 = 0$, in which case $g$ includes a point
mass at zero) to very large. Maximizing the likelihood $L(g)$ then
becomes a convex optimization problem in $\pi$ which can be solved
efficiently using interior point methods \citep{koenker2014convex},
sequential quadratic programming methods \citep{mixsqp}, or, more
simply---though less efficiently for large problems---using
accelerated EM algorithms \citep{Varadhan2008}. The conditional
distributions $p(\theta_j \| \bx, \bs, \hat{g})$ are analytically
tractable, and the posterior mean $E(\theta_j \| \bx, \bs, \hat{g})$
provides a shrinkage point estimate for $\theta_j$. See
\citet{stephens2017false} for details and various embellishments,
including generalizing the normal likelihood to a $t$ likelihood.

The representation \eqref{eqn:gnorm} provides a flexible family of
unimodal and symmetric distributions. Indeed, with a sufficiently
large and dense grid $\omega_0, \ldots, \omega_K$, the distribution $g$
in \eqref{eqn:gnorm} can arbitrarily accurately approximate any scale
mixture of normals. This family includes, as a special case, the
distributions used in
\citet{Clyde2000Flexible}, \citet{Johnstone2005Empirical}, and many
others \citep[e.g., the Horseshoe prior of][]{carvalho2010horseshoe}.
In this sense, \ash{} is more flexible than these existing EB
approaches. Furthermore, in many ways this more flexible approach
actually {\it simplifies} inference; by fixing the $\omega_k$ on a
dense grid, maximizing the likelihood $L(g)$ becomes a convex
optimization problem.
 
It is possible to implement EB methods for even broader families
$\G$. Indeed, \citet{koenker2014convex}, \citet{koenker2015rebayes}
provide methods and software for a fully non-parametric solution; {\em
  i.e.,} $\G$ is the set of all distributions on the real line.
However, the resulting maximum likelihood estimate $\hat{g}$ is then
discrete, which in the setting we consider here is unrealistic. More
generally, in many settings---including those considered
here---shrinkage towards zero is a desired outcome, and restricting
$\G$ to distributions that are unimodal at zero seems an attractive
and flexible way to achieve this.

\subsection{Signal Denoising via EB Shrinkage} \label{sec:denoise}

Here we introduce the homoskedastic Gaussian non-parametric regression
problem and summarize how it can be solved using the EB shrinkage
methods as in \citet{Johnstone2005Empirical}.

The homoskedastic Gaussian non-parametric regression problem has
essentially the same structure as the homoskedastic version of the
normal means problem (eq.~\ref{eqn:normalmeans}), but with the crucial
difference that the means to be estimated, denoted by $\bmu = (\mu_1,
\ldots, \mu_T)^{\top}$, are expected to spatially structured.  By
spatially structured, we mean that $\mu_t$ will often be similar to
$\mu_{t^{*}}$ for small $|t - t^{*}|$, though we do not rule out
occasional abrupt changes in $\bmu$.

In other words, homoskedastic Gaussian non-parametric regression
involves estimating a spatially structured mean $\bmu = (\mu_1,
\ldots, \mu_T)^{\top}$ from Gaussian observations $\bY = (y_1, \ldots,
y_T)^{\top}$ with standard error $\sigma$,
\begin{equation}
\label{eqn:nonparam-homo}
\bY \,|\, \bmu \sim N_T(\bmu,\sigma^2 I_T),
\end{equation}
where $I_T$ is the $T \times T$ identity matrix. Here, $t = 1, \ldots,
T$ indexes location in a one-dimensional space, such as time or, as in
a later example, location along the genome. For convenience, we assume
$T = 2^J$ for some integer $J$, as is common in multi-scale analyses.

Although the assumption that $\bmu$ is spatially structured is very
different from the sparsity assumption made by the EB shrinkage
methods described above, EB shrinkage methods can nonetheless be used
to solve this non-parametric regression problem
\citep{Johnstone2005Empirical}. The key idea is to apply a discrete
wavelet transform (DWT) to \eqref{eqn:nonparam-homo}. The DWT can be
expressed using an orthogonal $T \times T$ matrix $W$ that depends on
the wavelet basis chosen. Pre-multiplying \eqref{eqn:nonparam-homo} by
$W$ yields
\begin{equation}  
W\bY \,|\, W\bmu \sim N_T(W\bmu, \sigma^2 WW^{\top}).
\end{equation}
Note that $WW^{\top}=I_T$, so we write this as
\begin{equation}
\label{eqn:wc_homo}
\tbY \,|\, \tbmu \sim N_T(\tbmu, \sigma^2 I_T),
\end{equation}
in which $\tbY \colonequals W \bY = (\tilde{y}_1, \ldots,
\tilde{y}_T)^{\top}$ are the empirical wavelet coefficients (WCs), and
$\tbmu \colonequals W \bmu = (\tilde{\mu}_1, \ldots,
\tilde{\mu}_T)^{\top}$ are the (unknown) wavelet coefficients to be
estimated.

A key feature of the DWT is that if $\bmu$ is spatially structured,
many of the wavelet coefficients $\tbmu$ will be close to zero, and
vice versa \citep{mallat}. Thus, the DWT has changed the problem from
fitting \eqref{eqn:nonparam-homo} under the assumption that $\bmu$
spatially structured to fitting \eqref{eqn:wc_homo} under the
assumption that many of the WCs $\tbmu$ will be close to zero
\citep{donoho95}. This is easily achieved by the sparsity-inducing EB
shrinkage methods described above; it simply requires setting $\bx =
\tbY$, $\btheta = \tbmu$, $s_j^2 = \sigma^2$, and choosing $\G$ to
capture the assumption that $g$ has most of its mass near zero. The
value of $\sigma$ is of course typically unknown, but it can be
estimated by a number of simple methods \citep[e.g.,~equation 2 or 3
  from][]{Brown2007Variance}. In practice, it is important to group
the WCs by their resolution level before shrinking; see the note
below.

The EB procedure yields shrinkage estimates, $\hat{\tbmu}$, of the WCs
$\tbmu$, which can be reverse-transformed to obtain estimates for
$\bmu$:
\begin{equation}
\label{eqn:rev}
\hat{\bmu} \colonequals W^{-1} \hat{\tbmu} = W^T \hat{\tbmu}.
\end{equation}
This outlines the basic strategy used by
\citet{Johnstone2005Empirical} and implemented in the R software
package {\tt EbayesThresh} \citep{johnstone2005ebayesthresh}. 
\section{Methods}

Here, we extend the ideas from \citet{Johnstone2005Empirical} for the
homoskedastic Gaussian case and apply them to more general signal
denoising problems. First, we consider Gaussian data with
spatially structured mean {\it and} spatially structured variance
(Section~\ref{sec:het-gaussian-data}). In this setting, our methods
provide estimates for both the mean and variance. Second, we consider
denoising Poisson data (Section~\ref{sec:poisson-data}). In this setting,
the variance depends on the mean, so a spatially structured mean
implies spatially structured variance. Both settings require shrinkage
methods that can deal with heteroskedastic errors; we use the \ash{}
method from \citet{stephens2017false}. We call these methods
\mysmash{}, an abbreviation of ``SMoothing by Adaptive SHrinkage.''

\subsection{Heteroskedastic Gaussian Data}
\label{sec:het-gaussian-data}

The heteroskedastic analog of \eqref{eqn:nonparam-homo} is
\begin{equation}
\label{eqn:nonparam}
\bY \,|\, \bmu \sim N_T(\bmu,D),
\end{equation}
where $D$ is the diagonal matrix with diagonal entries $\bsigma^2 =
(\sigma_1^2, \ldots, \sigma_T^2)$.

Our goal here is to fit \eqref{eqn:nonparam} when both $\bmu$ and
$\bsigma^2$ are spatially structured. We consider, in turn, (i)
estimating $\bmu$ when $\bsigma^2$ is known, (ii) estimating
$\bsigma^2$ when $\bmu$ is known, and (iii) estimating $\bmu$ and
$\bsigma^2$ when both are unknown.

\subsubsection{Estimating $\mu$ with $\sigma^2$ Known} 
\label{sec:estimate-mu-given-sigma}

As in the homoskedastic case, the first step is to transform
\eqref{eqn:nonparam} using a wavelet transform,
\begin{equation}  
W\bY \,|\, W\bmu \sim N_T(W\bmu, WDW^{\top}),
\end{equation}
which we write as
\begin{equation} \label{eqn:wc}
\tbY \,|\, \tbmu \sim N_T(\tbmu, WDW^{\top}).
\end{equation}
As before, the $\tbY \colonequals W \bY = (\tilde{y}_1, \ldots,
\tilde{y}_T)^{\top}$ are the empirical WCs, and the $\tbmu
\colonequals W \bmu = (\tilde{\mu}_1, \ldots, \tilde{\mu}_T)^{\top}$
are the unknown WCs to be estimated. Unlike the homoskedastic case,
the covariance matrix of the empirical WCs in \eqref{eqn:wc} is no
longer diagonal and, in particular, the diagonal entries ({\em i.e.,}
the variances) are no longer the same.

To account for different variances among the WCs, we apply EB
shrinkage to the marginal distributions from \eqref{eqn:wc},
\begin{equation}
\label{eqn:marginal}
\tY_j \,|\, \tmu_j  \sim N(\tmu_j, \omega^2_j),
\end{equation}
in which
\begin{equation}
\omega^2_j = \sum_{t=1}^T \s^2_t W_{jt}^2, \qquad j = 1,\ldots,T.
\label{eqn:marginal-variance}  
\end{equation}
Specifically, to obtain the estimate $\hat{\tbmu}$, we apply \ash{}
(Section~\ref{sec:EB}), which fits a large mixture of unimodal
distributions, $g$, to the data, $x_j = \tY_j, s_j^2 = \omega_j^2$ $(j
= 1, \ldots, T)$. As in the homoskedastic case
(Section~\ref{sec:denoise}), applying EB shrinkage to the WCs yields
posterior mean estimates $\hat{\tbmu}$, from which estimates
$\hat{\bmu}$ are obtained by inverting the wavelet transform
(eq.~\ref{eqn:rev}).  Although this strategy accounts for
heteroskedacity in the WCs, it ignores correlations among them. We are
not alone in making this simplification; see
\citet{Silverman1999Wavelets} for example.

The simple but crucial point here is that the shrinkage step requires
EB methods that can solve the normal means problem with
heteroskedastic variances. Most treatments of the normal means
problem (including {\tt EbayesThresh}) avoid this complication,
whereas \ash{} is well suited to handling this situation. 

\subsubsection{Estimating $\sigma^2$ with $\mu$ Known}

To estimate the variance $\bsigma^2 = (\sigma_1^2, \ldots,
\sigma_T^2)$, we apply wavelet shrinkage methods to the squared
deviations from the mean, similar to the approaches of
\citet{Delouille2004Smooth} and \citet{Cai2008Adaptive}.
Specifically, we define
\begin{eqnarray}
\label{eqn:varobs1}
Z_t^2 \colonequals (y_t - \mu_t)^2,
\end{eqnarray}
and note that $\EE(Z_t^2) = \s_t^2$, so that estimating $\bsigma^2$
reduces to a mean estimation problem with ``observations'' $\bZ^2
\colonequals (Z_1^2, \ldots, Z_T^2)$.

As in the procedure for estimating $\bmu$ given $\bm{\sigma}^2$
(Section~\ref{sec:estimate-mu-given-sigma}), we estimate $\bm{\sigma}^2$
by fitting the \ash{} model (Section~\ref{sec:EB}) to the observations
$x_t = Z_t^2$. To apply \ash{}, we need an estimate of the variance of
each $Z_t^2$. We use $s_t^2 = \frac{2}{3}Z_t^4$, which is an unbiased
estimator of the variance. (If $Z^2 \sim \s^2 \chi_1^2$, then
$\EE(Z^4) = 3\s^4$ and $\mathrm{Var}(Z^2) = 2\s^4$.)

This approach effectively approximates the wavelet-transformed values
$\tilde{\bZ}^2 \colonequals W\bZ^2 = (\tilde{Z}_1^2, \ldots,
\tilde{Z}_T^2)^{\top}$ by a Gaussian distribution when really they are
linear combinations of $\chi_1^2$ random variables. Despite this
approximation, we have found this procedure to work well in practice
in most cases, perhaps with a tendency to oversmooth quickly-varying
variance functions.

\subsubsection{Estimating $\mu$ and $\sigma^2$ Jointly} 

To deal with the (more common) case where both mean and variance are
unknown, we simply iterate the above procedures. That is, the
algorithm consists of repeating the following two steps:
\begin{enumerate}
  
\item Estimate $\bm{\mu}$ as if $\bm{\s}^2$ is known (with $\bsigma^2$
  set to the estimate $\hat{\bsigma}^2$ obtained from the previous
  iteration).
  
\item Estimate $\bm{\s}^2$ as if $\bm{\mu}$ is known (with $\bmu$ set
  to the estimate $\hat{\bmu}^2$ obtained from Step 1).

\end{enumerate}
To initialize the algorithm, we estimate the variance $\bm{\s}^2$ as
\begin{equation*}
\label{eqn:initial_var_est}
\hat{\s}_t^2 = \frac{1}{2}\left((y_t-y_{t-1})^2+(y_t-y_{t+1})^2\right),
(t = 1,\ldots,T),
\end{equation*}
defining $y_0 = y_n$ and $y_{T+1} = y_1$ (equivalent to putting the
observations on a circle). 
  
We cannot guarantee that this procedure will converge, but in our
simulations we found that two iterations of steps 1 and 2 reliably
yielded accurate results. (So the full procedure consists of
initialization, running steps 1 and 2, then running steps 1 and 2 a
second time.)

\subsection{Poisson Data}
\label{sec:poisson-data}

\def\Poi{\text{Pois}}
\def\Bin{\text{Bin}}

Now we consider estimating a spatially structured mean $\bmu = (\mu_1,
\ldots, \mu_T)^{\top}$ from Poisson data:
\begin{equation*}
y_t \sim \Poi(\mu_t), \qquad (t=1,\ldots,T).
\end{equation*}
For Poisson data, the analogue of the DWT is provided by the Poisson
multiscale models from \citet{Kolaczyk1999Bayesian,
  Timmermann1999Multiscale, Nowak2000Statistical}.  In brief, we
estimate $\bmu$ by applying \ash{} to shrink parameters within these
multi-scale models.

To motivate this approach, first recall the following elementary
distributional result: if $y_1$ and $y_2$ are independent, with $y_t
\sim \Poi(\mu_t)$ then
\begin{align*}
y_1 + y_2 & \sim \Poi(\mu_1 + \mu_2) \\
y_1 \,|\, (y_1 + y_2) & \sim \Bin(y_1+y_2, \mu_1/(\mu_1+\mu_2)).
\end{align*} 
To extend this to $T = 2 \times 2 = 4$, we introduce notation
$v_{i:j}$ to denote the sum $v_{i:j} = \sum_{t = i}^j v_t$ for vector
$v$. Then we have that
\begin{align} 
y_{1:4} & \sim \Poi(\mu_{1:4}) \label{eqn:y14} \\
y_{1:2} \,|\, y_{1:4} & \sim \Bin(y_{1:4}, \mu_{1:2}/\mu_{1:4})
\label{eqn:y12} \\
y_1 \,|\, y_{1:2} & \sim \Bin(y_{1:2}, \mu_1/\mu_{1:2}) \label{eqn:y1} \\ 
y_3 \,|\, y_{3:4} & \sim \Bin(y_{3:4}, \mu_3/\mu_{3:4}). \label{eqn:y3}
\end{align} 
Together, these models are equivalent to $y_t \sim \Poi(\mu_t)$, for
$t = 1, \ldots, 4$, and they decompose the overall distribution $y_1,
\ldots, y_4$ into parts involving aspects of the data at increasing
resolution; eq.~\ref{eqn:y14} represents the coarsest resolution (the
sum of all the data points), whereas (\ref{eqn:y1}, \ref{eqn:y3})
represent the finest resolution, and \eqref{eqn:y12} is the in-between
resolution. This representation suggests a reparameterization, from
$(\mu_1, \mu_2, \mu_3, \mu_4)$ to $(\mu_{1:4}, \bm{p})$, where
binomial parameters $\bp = (p_1, p_2, p_3) = (\mu_{1:2}/\mu_{1:4},
\mu_1/\mu_{1:2}, \mu_3/\mu_{3:4})$ control lower ($p_1$) and higher
resolution $(p_2, p_3)$ changes in the mean vector $\mu$.

This idea extends naturally to $T=2^J$ for any $J$, reparameterizing
$\bmu$ into its sum $\mu_{1:T}$ and the $T-1$ binomial probabilities
$\bm{p} = (p_1, \ldots, p_{T-1})$ that capture features of $\bmu$ at
different resolutions. This can be viewed as the Poisson analogue of the
Haar wavelet transform.

In this reparameterization, $p_j = \frac{1}{2}$ for all $j = 1,
\ldots, T-1$ corresponds to the case of a constant mean vector, and
values of $p_j$ far from $\frac{1}{2}$ correspond to large changes in
$\mu$ (at some scales). Therefore, estimating a spatially structured
$\bmu$ can be achieved by shrinkage estimation of $\bp$, with
shrinkage towards $p_j = \frac{1}{2}$. Both
\citet{Kolaczyk1999Bayesian} and \citet{Timmermann1999Multiscale} use
dedicated Bayesian models to achieve this shrinkage by introducing a
prior distribution on elements of $\bp$ that is a mixture of a point
mass at $\frac{1}{2}$ (resulting in shrinkage toward $\frac{1}{2}$)
and a Beta distribution. We take a different approach,
reparameterizing the $p_j$'s as $\alpha_j =
\log\big(\frac{p_j}{1-p_j}\big)$, $j = 1, \ldots, T-1$, then using
\ash{} to shrink the parameters $\alpha_j$ towards zero, since
$\alpha_j = 0$ when $p_j = \frac{1}{2}$. Since \ash{} is based on
solving the normal-means problem, this is effectively making a normal
approximation to the likelihood for the parameters $\alpha_j$ (this is
not the same as making a normal approximation for the data).

To obtain a normal approximation to the likelihood for $\bm{\alpha} =
(\alpha_1, \ldots, \alpha_{T-1})$, it suffices to have an estimate
$\hat{\alpha}_{j}$ and corresponding standard error $\hat{s}_j$ for
each $j$. This problem---estimating a
log-odds ratio and its standard error---has been well studied
\citep[e.g.,][]{Gart1967Bias}. The main challenge is in dealing
satisfactorily with cases where the maximum likelihood estimator for
$\alpha_j$ is infinite. We use estimates based on results from
\citet{Gart1967Bias}; see Appendix~\ref{app:reconstruction}.

Applying \ash{} to the estimates $\hat{\alpha}_j$ and standard errors
$\hat{s}_j$ yields a posterior distribution for each $\alpha_j$. The
simplest way to convert this to an estimate of the mean, $\bmu$, is to
estimate $\alpha_j$ by its posterior mean, then reverse the above
reparameterization. (Recovering $\mu$ also requires an estimate
$\mu_{1:T}$---we take its maximum-likelihood estimate, which is $y_1 +
\cdots + y_T$.) The resulting estimate of each $\mu_t$ is the
exponential of the posterior mean for $\log\mu_t$ (because each
$\log\mu_t$ is a linear combination of the $\alpha_j$'s). Alternatively,
we can estimate each $\mu_t$ by approximating its posterior mean using
the delta method; see Appendix \ref{app:reconstruction}. Both methods
are implemented in our software. For the results below, we use the
delta method because it is more comparable with previous approaches
that estimate $\bmu$ on the original scale rather than the logarithmic
scale.

\subsection{Practical implementation details} 

In practice, we follow these additional steps, guided by prior
work, to improve performance and reduce effort:
\begin{itemize}

\item Rather than use a single wavelet transform, we use the
  ``translation invariant'' wavelet transform (also called the
  ``non-decimated'' wavelet transform), which averages results over
  all $T$ possible rotations of the data (effectively treating the
  observations as coming from a circle, rather than a line). Although
  not always necessary, this is a standard trick to reduce artifacts
  that can occur near discontinuities in the underlying signal, and
  can often improve performance
  \citep[e.g.,][]{Coifman1995Translationinvariant}. Implementation of
  the translation invariant wavelet transform for the Poisson model is
  described in Appendix~\ref{app:TI}.
  
\item The non-decimated wavelet transform yields $T$ WCs at each of
  the $J = \log_2(T)$ resolution levels. We follow
  \citet{Johnstone2005Empirical} in applying EB shrinkage separately
  to the $T$ WCs at each resolution level, so that a different
  distribution $g$ is estimated at each resolution. This is important
  because sparsity in the WCs $\tilde{\mu}_j$ will likely vary with
  resolution, and therefore the amount of shrinkage to apply should
  also be resolution-specific.
  
\item Although we have presented the DWT as a matrix-vector
  multiplication, which would naively take $O(T^2)$ operations, in
  practice there exist more efficient algorithms taking only
  $O(T\log_2 T)$ operations \citep{Beylkin1992Ontherepresentation,
    Coifman1995Translationinvariant}. These are implemented in the R
  package {\tt wavethresh} \citep{Nason_wavethresh}, for example.
  
\end{itemize}

\section{Results}

We conducted a wide range of numerical experiments to compare
\mysmash{} against the existing methods for wavelet-based signal
denoising. Before presenting the results from these experiments in
Section~\ref{sec:simulations}, we first illustrate the features of
\mysmash{} in a small example (Section~\ref{sec:illustration}). In
Section~\ref{sec:applications}, we show two applications of \mysmash{}.

We have developed a companion repository, available at
\url{https://github.com/stephenslab/smash-paper}, containing all the
source code (R and MATLAB) and data used to generate the results,
figures and tables shown here. This resource includes a ``Shiny'' Web
app \citep{Shiny} for browsing the full results of the the simulation
study in Section~\ref{sec:gauss-mean-est}.

\subsection{Illustration}
\label{sec:illustration}

\begin{figure}[p]
\centering
\includegraphics[width=5.2in]{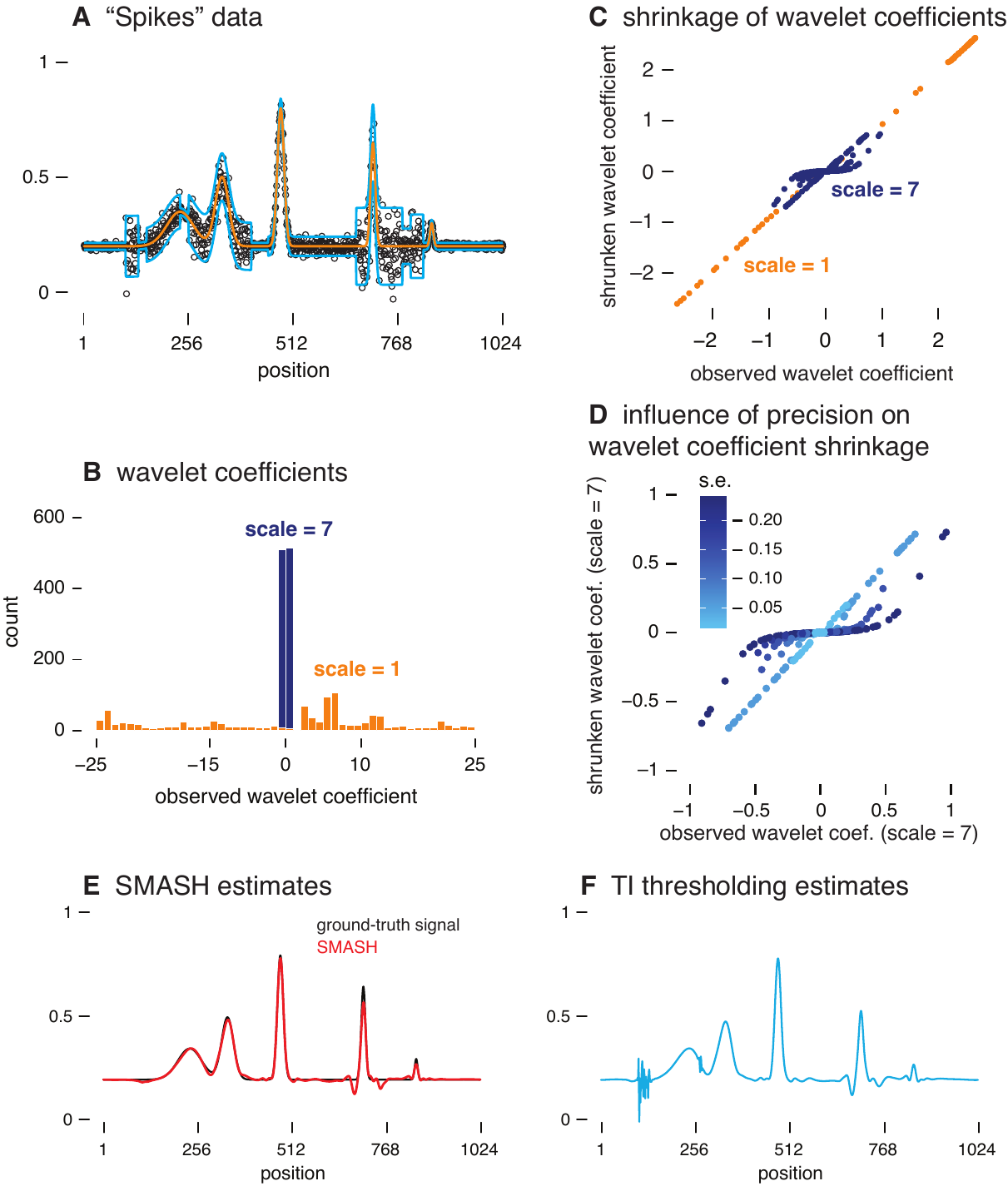}
\caption{Illustration of our denoising method, \mysmash{}, based on
  the EB shrinkage method, \ash{}. Panel A shows the ``Spikes'' mean
  function (\textcolor{darkorange}{orange line}), with $\pm 2$
  standard deviations given by the ``Clipped Blocks'' function
  (\textcolor{dodgerblue}{light blue lines}). The simulated data
  $\bm{y} = (y_1, \ldots, y_T)^{\top}$ are shown as {\bf black
    circles} ($\circ$). Panel B contrasts the distributions of the
  simulated wavelet coefficients (WCs), $\tilde{y}_j$, at a coarser
  scale (scale = 1, \textcolor{darkorange}{orange}) and at a finer
  scale (scale = 7, \textcolor{darkblue}{dark blue}). Note that the
  scale = 7 WCs are much more concentrated near zero because the
  signal is smoother at this finer scale. Panel C contrasts the \ash{}
  shrinkage at these two scales; the scale = 7 WCs are strongly shrunk
  toward zero, whereas the scale = 1 WCs are not shrunk nearly as
  much. In this case, \ash{} infers that the scale = 7 WCs are heavily
  concentrated around zero, and consequently \ash{} shrinks them more
  strongly. Panel D illustrates that \ash{} shrinks WCs differently
  depending on their precision; specifically, WCs that are less
  precise---{\em i.e.}, higher standard error (s.e.)---are shrunk more
  strongly toward zero. Panels E, F show the signals, $\bmu = (\mu_1,
  \ldots, \mu_T)^{\top}$, reconstructed by \mysmash{}
  (\textcolor{orangered}{red}) and translation-invariant (TI)
  thresholding (\citealt{Gao1997Wavelet}; \textcolor{dodgerblue}{light
    blue}); compare against the true mean function ({\bf black}). The
  TI thresholding estimate shows notable artifacts. This example is
  implemented by the ``Spikes'' demo in the companion source code
  repository.}
\label{fig:simple_eg}
\end{figure}

Figure~\ref{fig:simple_eg} illustrates the key features of \mysmash{}
applied to smoothing a heteroskedastic Gaussian signal. The data in
this example were simulated with a mean and variance that are both
spatially structured (Figure~\ref{fig:simple_eg}, Panel A).

The first step of \mysmash{} is to compute the WCs at different scales
by applying the DWT. Each observed wavelet coefficient, $\tilde{y}_j$,
can be viewed as a noisy estimate of some unknown ``true'' wavelet
coefficient, $\tilde{\mu}_j$. These wavelet coefficients
$\tilde{\mu}_j$ will be estimated using Empirical Bayes shrinkage
(eq.~\ref{eqn:marginal}). Each WC, $\tilde{y}_j$, is also associated
with a standard error, $\omega_j^2$, that depends on the simulated
variance of the data (eq.~\ref{eqn:marginal-variance}).

A key idea behind wavelet denoising is to ``shrink'' the observed WCs
towards zero, resulting in an estimate of the mean that is smoother
than if it were based solely on the observed data. A crucial question
is, of course, how much to shrink. The \ash{} shrinkage method, which
underlies \mysmash{}, adapts the amount of shrinkage to the data in
two distinct ways.

If many observed WCs $\tilde{y}_j$ are ``large'' at a particular scale
(compared with their standard errors), \ash{} infers that, at this
scale, many of the true WCs $\tilde{\mu}_j$ must also be large---that
is, the estimated distribution $g$
(\ref{eqn:ebshrink}--\ref{eqn:ebshrink2}) has a long
tail. Consequently, \ash{} shrinks less at this scale than at scales
where few observed WCs are large, in which case the estimated $g$ will
have a short tail. This is illustrated in Figure~\ref{fig:simple_eg},
Panels B and C; at scale = 1, many observed WCs are large (Panel B),
so very little shrinkage is applied to these estimates (Panel C). By
contrast, at scale = 7, few observed WCs are large (Panel B), and
therefore stronger shrinkage is applied (Panel C). This adaptive
feature is also characteristic of other EB shrinkage methods, but the
family of unimodal distributions underlying \ash{} is more flexible
than other methods, increasing its potential to adapt to different
contexts.

Second, because the posterior distribution \eqref{eqn:post}
incorporates the standard error of each observation, shrinkage is
adaptive to the standard error; at a given scale, WCs $\tilde{y}_j$
with larger standard errors $\omega_j$ are shrunk more strongly than
WCs with small standard errors. (In this example, the standard errors
vary among WCs due to the spatially structured variance of the
simulated data.) This is illustrated in Panel D.

The end result is that (i) data that are consistent with a smooth
signal are smoothed more strongly, and (ii) smoothing is stronger in
areas of the signal with greater variance. The smoothed signal from
\mysmash{} (Figure~\ref{fig:simple_eg}, Panel E) is noticeably more
accurate than the signal estimated using TI thresholding in Panel F
(in which the variance is estimated using the ``median absolute
deviation,'' or RMAD, method of \citealt{Gao1997Wavelet}).

We revisit this simulation scenario in
Section~\ref{sec:gauss-mean-est}, where we compare the performance of
\mysmash{} against signal denoising methods in many simulated data
sets.

\subsection{Simulations}
\label{sec:simulations}

We investigated the signal denoising performance of \mysmash{} against
existing approaches in data sets simulated from Gaussian and Poisson
distributions.

\subsubsection{Gaussian Mean Estimation}
\label{sec:gauss-mean-est}

\begin{figure}[t!]
\centering
\includegraphics[width=4.75in]{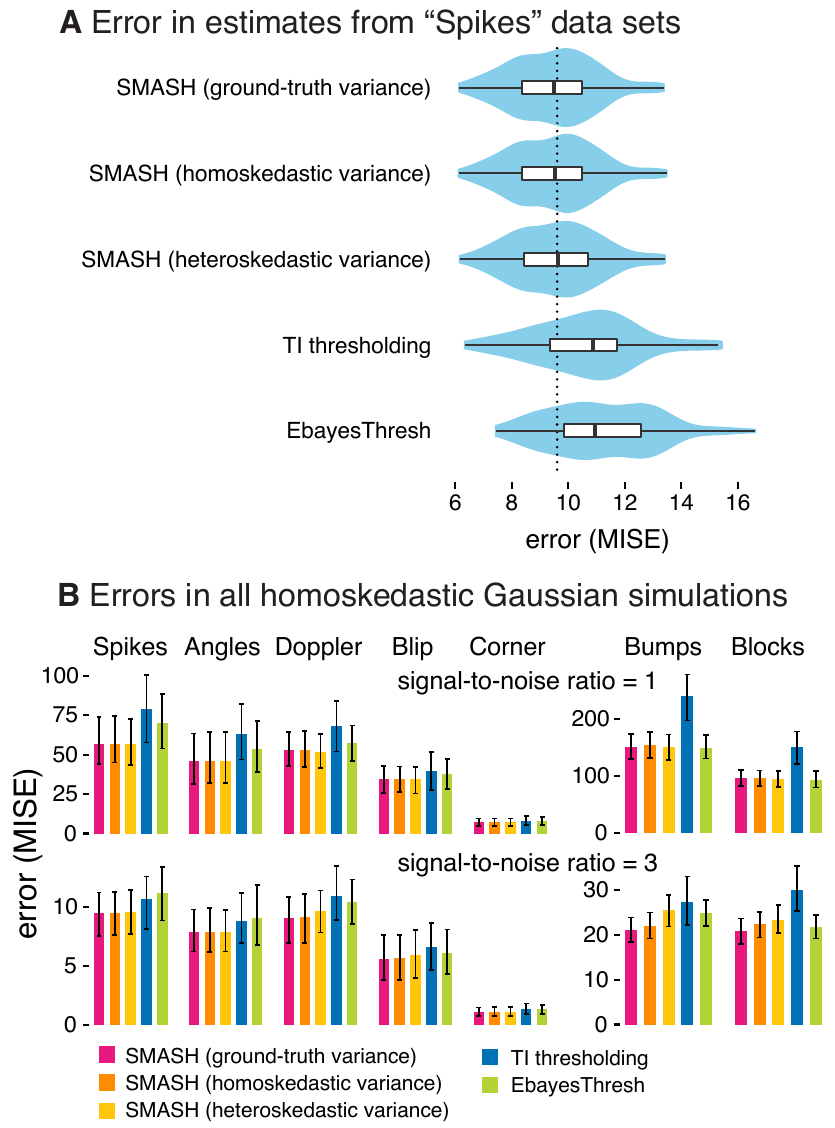}
\caption{Accuracy of mean signal estimates in data sets simulated with
  homoskedastic Gaussian noise. Panel A shows violin plots (and inset
  boxplots) summarizing the error (MISE) of the estimates in the
  ``Spikes'' simulation scenario with constant variance and a
  signal-to-noise ratio of 3. In Panel B, bars give
    the average error (MISE) in the mean estimates across all
    simulations; error bars show the 10\% and 90\%
    quantiles. Functions used to simulate data sets in each scenario
    (Panel B columns) are shown in Figure
    \ref{fig:gaussian_mean_signals}. Methods compared are: \mysmash{}
  with homoskedastic variances; \mysmash{} allowing for
  heteroskedastic variances; \mysmash{} when the ground-truth variance
  is provided; TI thresholding; and EbayesThresh. (Both TI
  thresholding and EbayesThresh assume homoskedastic variances.) In
  the ``Spikes'' scenario (Panel A), all variants of \mysmash{}
  outperformed TI thresholding and Ebayesthresh; overall (Panel B),
  \mysmash{} consistently performed as well or better than the other
  methods.}
    \label{fig:gaus_homo}
\end{figure}

In our first set of simulations, we ran different methods for
estimating a spatially structured mean from Gaussian-distributed
observations, and assessed accuracy of the estimates. Our simulation
study was modeled after \citet{Antoniadis2001Wavelet}. Specifically,
we used many of the same test functions (7 mean functions, 5 variance
functions) and two different signal-to-noise ratios, 1 and 3 (Figures
\ref{fig:gaussian_mean_signals} and
\ref{fig:gaussian_variance_signals}). For each combination of
simulation settings, we simulated 100 data sets, each with a signal of
length $T = \mbox{1,024}$, and applied the signal denoising methods to
each of the simulated data sets. In all cases, we ran three variations
of \mysmash{}: when the variance function was estimated, allowing for
heteroskedasticity; when variance was estimated, assuming
homoskedasticity; and when \mysmash{} was provided with the
ground-truth variance function, which could be viewed as a ``gold
standard.'' We compared these \mysmash{} variants against the
Translation Invariant (TI) thresholding method
\citep{Coifman1995Translationinvariant}, which was one of the methods
shown to performing best in \citet{Antoniadis2001Wavelet}. We also
compared against the Empirical Bayes shrinkage procedure,
``EbayesThresh'' \citep{johnstone2005ebayesthresh}.  For all results
shown in the figures and tables below, the methods used the Symmlet8
wavelet basis \citep{Daubechies1992Ten}. To assess performance of the
methods, we report the mean integrated squared error (MISE), which
summarizes the difference between the ground-truth signal and the
estimated mean signal \citep{nason-1996}. R and MATLAB scripts
implementing these comparisons, as well as the results generated using
these scripts, are provided in the companion repository.

We focus initially on the simulations with homoskedastic
variance. Figure \ref{fig:gaus_homo} compares the performance of each
of the methods in this setting. In the ``Spikes''
  scenario (Panel A), all three variants of \mysmash{} outperformed
  EbayesThresh and TI. Further, the three \mysmash{} variants yielded
  estimates of comparable accuracy. This illustrates that allowing for
  heteroskedasticity when the truth is homoskedastic can sometimes be
  done with little or no loss of accuracy. Most of the other
  simulation settings with homoskedastic variance show similar trends
  (Figure \ref{fig:gaus_homo}, Panel B). For the most difficult
  settings---``Bumps'' and ``Blocks'' with a signal-to-noise ratio of
  1---EbayesThresh achieved similar accuracy to \mysmash{}, whereas TI
  thresholding performed much worse.

\begin{figure}[t!]
\centering
\includegraphics[width=\textwidth]{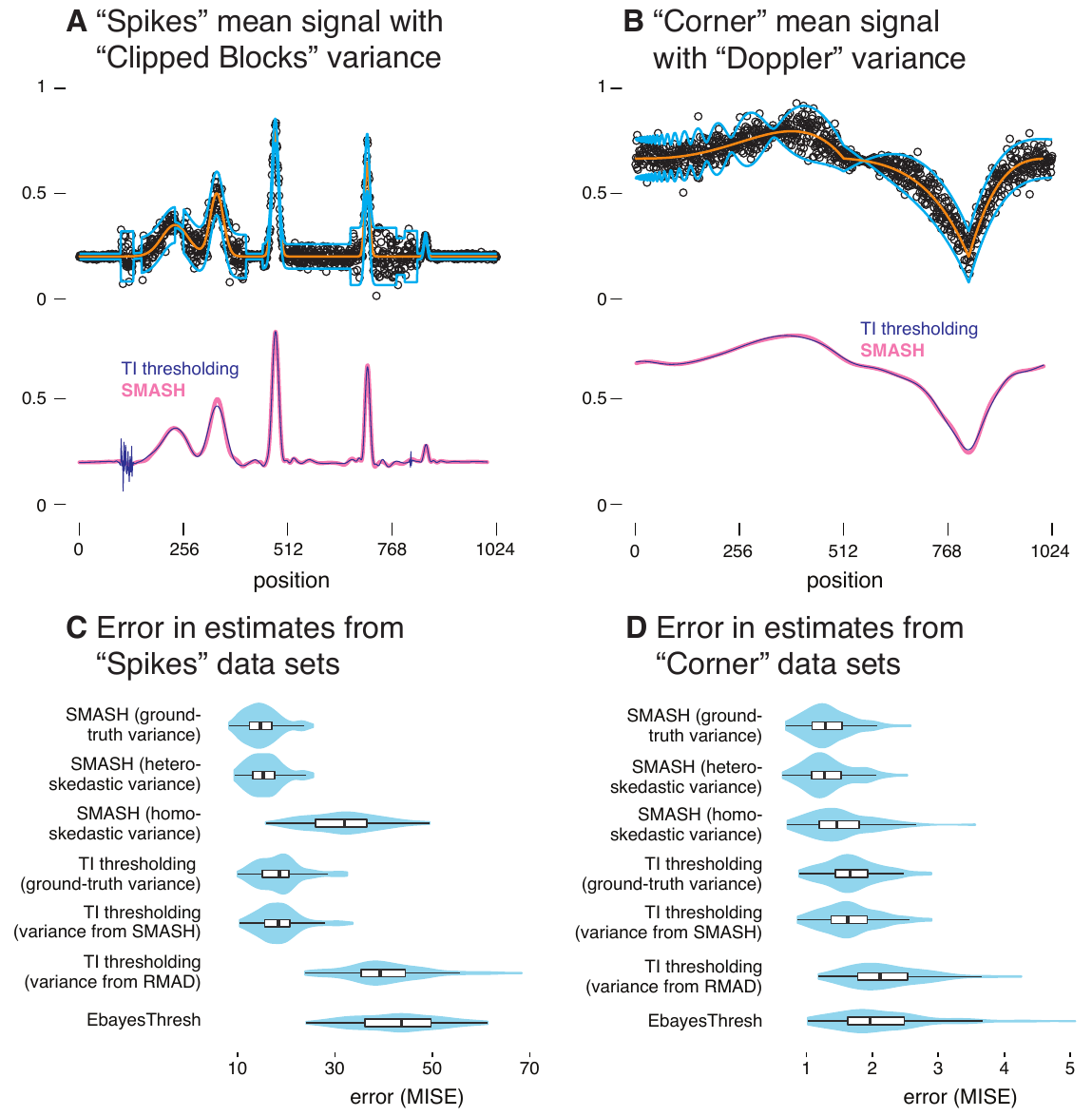}
\caption{Illustration of signal denoising methods in Gaussian data
  sets simulated with heteroskedastic errors. Panels A and B depict
  the mean signals (\textcolor{darkorange}{orange lines}) and variance
  functions (\textcolor{dodgerblue}{light blue lines}, showing $\pm 2$
  standard deviations) used to simulate the data. An example simulated
  data set is shown in each case ({\bf black circles}, $\circ$). The
  signals recovered by TI thresholding with RMAD variance estimates
  (\textcolor{darkblue}{dark blue line}) and \mysmash{} with
  estimated heteroskedastic variances (\textcolor{pink}{pink line})
  are also shown. Panels C and D are violin plots (and inset boxplots)
  summarizing the error (MISE) in the mean estimates. Methods compared
  are: \mysmash{} with homoskedastic variances, with the ground-truth
  variances, and allowing for heteroskedastic variances; TI
  thresholding with \mysmash-estimated variances, with RMAD-estimated
  variances, and with ground-truth variances; and EbayesThresh.}
\label{fig:gaus_hetero}
\end{figure}

\begin{figure}[t!]
\centering
\includegraphics[width=6in]{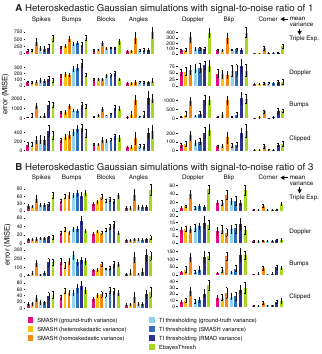}
\caption{Comparison of signal denoising methods in
    Gaussian data sets simulated with heteroskedastic error, with a
    signal-to-noise ratio of 1 (Panel A) and 3 (Panel B). Bars give
    the average error (MISE) in the mean estimates across all
    simulations; error bars show the 10\% and 90\% quantiles. Each
    scenario is defined by a combination of the mean function
    (columns) and variance function (rows) used to simulate the data
    (these functions are shown in Figures
    \ref{fig:gaussian_mean_signals} and
    \ref{fig:gaussian_variance_signals}). In each scenario, 100 data
    sets were simulated. Methods compared are: three variants of
    \mysmash{} (with homoskedastic variances, ground-truth variances,
    and allowing for heteroskedastic variances); three variants of TI
    thresholding (with \mysmash-estimated variances, RMAD-estimated
    variances, and ground-truth variances); and EbayesThresh.}
\label{fig:gaus_hetero_full}
\end{figure}

Next, we examine the performance of the same methods in simulated data
sets with heteroskedastic errors. Since the performance of the TI
thresholding method with homoskedastic variances was consistently poor
(see the interactive plot), we considered three different ways to
allow for heteroskedastic variances in TI thresholding: providing the
ground-truth variance; estimating the variances using \mysmash{}; and
estimating the variances using the extended RMAD method from
\cite{Gao1997Wavelet} (henceforth ``RMAD'' for short).

Figure \ref{fig:gaus_hetero} provides a detailed view of performance
on data sets simulated with a signal-to-noise ratio of 3: the
``Spikes'' mean function with the ``Clipped Blocks'' variance function
(Figure~\ref{fig:gaus_hetero}, Panels A, C); and the ``Corner'' mean
function with the ``Doppler'' variance function
(Figure~\ref{fig:gaus_hetero}, Panels B, D).
Figure~\ref{fig:gaus_hetero_full} summarizes the
  results from all simulations. The results of all these simulations
  can be explored interactively in the Shiny plot included in the
  companion repository.

  Allowing for heteroskedasticity in \mysmash{}
  substantially improved its accuracy in all settings (compare the
  yellow and orange bars in Figure~\ref{fig:gaus_hetero_full}).
  Further, in nearly all settings, \mysmash{} with estimated
  heteroskedastic variance generally performed at least as well as,
  and often much better than, EbayesThresh and all TI thresholding
  variants. While accuracy improvements were greatest in data sets
  simulated with sudden, large changes to the variance (``Bumps'' and
  ``Clipped'' variance functions), what is perhaps more remarkable is
  that \mysmash{} provided consistently competitive performance in all
  settings.

  We comment now on some other key trends in the results
  in Figure \ref{fig:gaus_hetero_full}. First, \mysmash{} with
  estimated heteroskedastic variance often achieved comparable
  accuracy to \mysmash{} with the ground-truth variance. However, some
  variance functions are harder to estimate than others (e.g., the
  ``Bumps'' and ``Blocks'' functions; see Figure
  \ref{fig:gaussian_variance_signals}), and in such cases providing
  the method with the ground-truth variance improved accuracy.
  Second, EbayesThresh generally performed much less competitively
  here than in the homoskedastic setting, which highlights the
  importance of accounting for heteroskedasticity. The most extreme
  example of this is in simulations with the ``Triple Exponential''
  variance test function, which has large changes in variance, but the
  changes are gradual enough that estimating the variance can be done
  accurately. Consistent with the results in Figure
  \ref{fig:gaus_homo}, \mysmash{} with homoskedastic variance
  consistently performed better than, or at least as well as,
  EbayesThresh.

  Finally, TI thresholding generally performed better
  when used with the \mysmash{} variance estimate than with the RMAD
  variance estimate. The largest differences in performance were in
  simulations with more abrupt changes to variances; indeed, the RMAD
  estimates performed well in simulations with the smoother ``Triple
  Exponential'' variance function. This suggests that the RMAD method
  works best in settings where the variance changes gradually.

\subsubsection{Gaussian Variance Estimation}
\label{sec:mfvb}

An unusual feature of \mysmash{} is that it performs joint mean and
variance estimation. We found no R packages for doing this in the
wavelet context. We only found one publication on wavelet-based
variance estimation, \cite{Cai2008Adaptive}, in which a wavelet
thresholding approach is applied to first-order differences in the
data.  Non-wavelet-based approaches related to this work include a
method by \cite{Fan1998Efficient}, which estimates the variance by
smoothing the squared residuals using local polynomial smoothing;
\cite{Brown2007Variance}, which uses difference-based kernel
estimators; and \cite{Menictas2015Variational}, which introduces a
Mean Field Variational Bayes (\mfvb{}) method for joint mean and
variance estimation. In all cases, we could not find publicly
available software implementations of these methods. However, we did
receive code implementing \mfvb{} via correspondence with M.~Menictas,
and we used this code in our comparisons.

\begin{figure}[t!]
\centering
\includegraphics[width=3in]{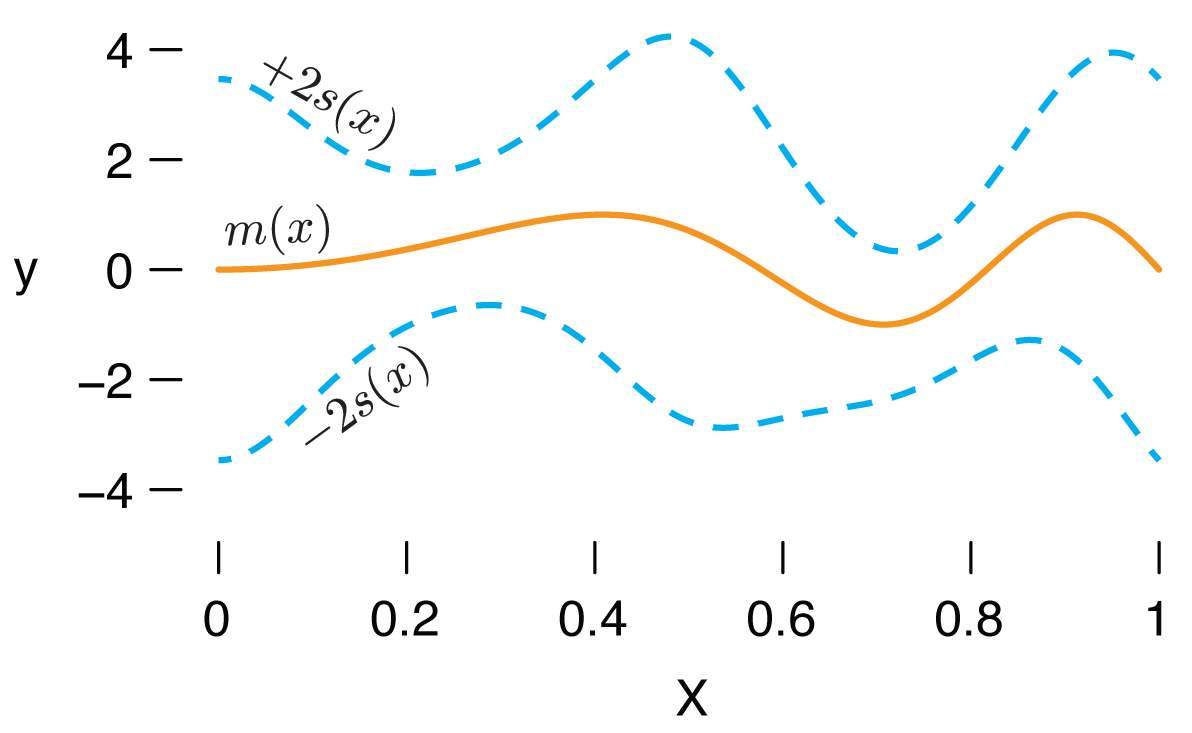}
\caption{The mean function, $m(x)$ (\textcolor{darkorange}{orange
    lines}), and $\pm 2$ standard deviations, $s(x)$
  (\textcolor{dodgerblue}{dashed, light blue lines}), used to simulate
  the data sets for comparing \mysmash{} and \mfvb{}. These are the
  same as the mean and standard deviation functions used in ``Scenario
  A,'' Figure 5 in \citet{Menictas2015Variational}.}
\label{fig:mfvb_fn}
\end{figure}

The \mfvb{} method is based on penalized splines, so it is not well
suited to many of the standard test functions in the wavelet
literature---these test functions often contain ``spiky'' local
features that are not well captured by splines. Therefore, to design a
fair comparison, we applied \mysmash{} and \mfvb{} to smooth mean and
variance functions; specifically, we generated data in the same way as
``Scenario A'' in Figure~5 from \citet{Menictas2015Variational} using
scripts kindly provided by M.~Menictas. The mean function and variance
function are shown in Figure~\ref{fig:mfvb_fn}.

We evaluated \mysmash{} and \mfvb{} in two scenarios. In the first
scenario, we simulated unevenly spaced data points: we independently
generated $T = 500$ pairs $(X_t, y_t)$, with $X_t \sim
\mathrm{Uniform}(0,1)$ and $y_t \,|\, X_t = x_t \sim
N(m(x_t),s(x_t)^2)$, in which $m(\,\cdot\,)$ and $s(\,\cdot\,)$ denote
the mean and standard deviation functions shown in
Figure~\ref{fig:mfvb_fn}. To assess accuracy, we computed the mean of
the squared errors (MSE) evaluated at 201 equally spaced points within
$[\min(X), \max(X)]$, where $\min(X)$ and $\max(X)$ are the smallest
and largest values of $X = (X_1, \ldots, X_T)$, respectively. We
computed the MSE separately for estimates of the mean and standard
deviation. For both \mysmash{} and \mfvb{}, estimates of the mean and
variance at each of the 201 equally spaced points were obtained by a
simple linear interpolation between the available estimates at the two
nearest flanking data points.

In this scenario, \mysmash{} could not be immediately applied to the
simulated data because the points were not equally spaced, and the
number of data points was not a power of 2. To address the first
issue, we followed the common practice of treating the observations as
if they were evenly spaced \citep[see][for discussion]{Sardy1999Wavelet}.

To deal with the second issue, we borrowed a standard trick used in
the wavelet literature; first, we reflected the data about the right
edge and extracted the first $2^{\lfloor\log_2(2T)\rfloor} = 512$ data
points, so that the number of data points in the new data set was a
power of 2, and so that the mean curve was continuous at the right
edge of the original data. Further, to ensure that the input to
\mysmash{} was periodic, we reflected the transformed data set about
its right edge, so that the final transformed signal was of length
1,024. After running \mysmash{}, the estimates of the mean and
variance functions were extracted from the first $T = 500$ positions.

In the second scenario, we simulated evenly spaced data points; we
independently generated $T = \mbox{1,024}$ pairs $(X_t,y_t)$, with the
$X_t$'s equally spaced on $[0,1]$. Performance was evaluated
separately for the mean and standard deviation as the mean of the MSEs
evaluated at each of the locations, $t = 1,\ldots,T$.

For each scenario, we simulated 100 data sets. These experiments are
implemented by in the ``Gaussian variance estimation'' analysis in the
companion repository.

\begin{table}[tb!]
\centering
\begin{tabular}{rcccc} 
& \multicolumn{2}{c}{\textbf{\textsf{Scenario 1}}} &
  \multicolumn{2}{c}{\textbf{\textsf{Scenario 2}}} \\
  \cmidrule(lr){2-3} \cmidrule(lr){4-5}
& \sf MSE (for mean) & \sf MSE (for s.d.) & \sf MSE (for mean)
  & \sf MSE (for s.d.) \\ 
\sf MFVB & \sf 0.0330 & \sf 0.0199 & \sf 0.0172 & \sf 0.0085 \\
\sf SMASH & \sf 0.0334 & \sf 0.0187 & \sf 0.0158 & \sf 0.0065 \\
\end{tabular}
\caption{Accuracy of \mysmash{} and MFVB in two simulation
  scenarios. In each simulation, accuracy is measured using the mean
  of squared errors (MSE). The table shows the MSE averaged over the
  100 simulations in each of the scenarios. The true mean and standard
  deviation (s.d.) functions are shown in Figure~\ref{fig:mfvb_fn}. In
  Scenario 1, the data are not equally spaced, and the number of data
  points is not a power of 2; in this setting, \mysmash{} is more
  accurate in estimating both the mean and s.d. In Scenario 2, the
  data are equally spaced, and the number of data points is a power of
  2; \mysmash{} again outperforms MFVB in both mean and
  s.d. estimation.}
\label{table:mfvb_comp}
\end{table}

Table \ref{table:mfvb_comp} shows, for each scenario, the mean error
(MSE) in the estimated mean and standard deviation, averaged over the
100 independent simulations. Despite the fact that these simulation
scenarios, particularly Scenario 1, seem better suited to \mfvb{} than
\mysmash{}, \mysmash{} performs comparably or better than \mfvb{} for
both mean and variance estimation.

\subsubsection{Poisson Data}

\begin{figure}[t!]
\centering
\includegraphics[width=5in]{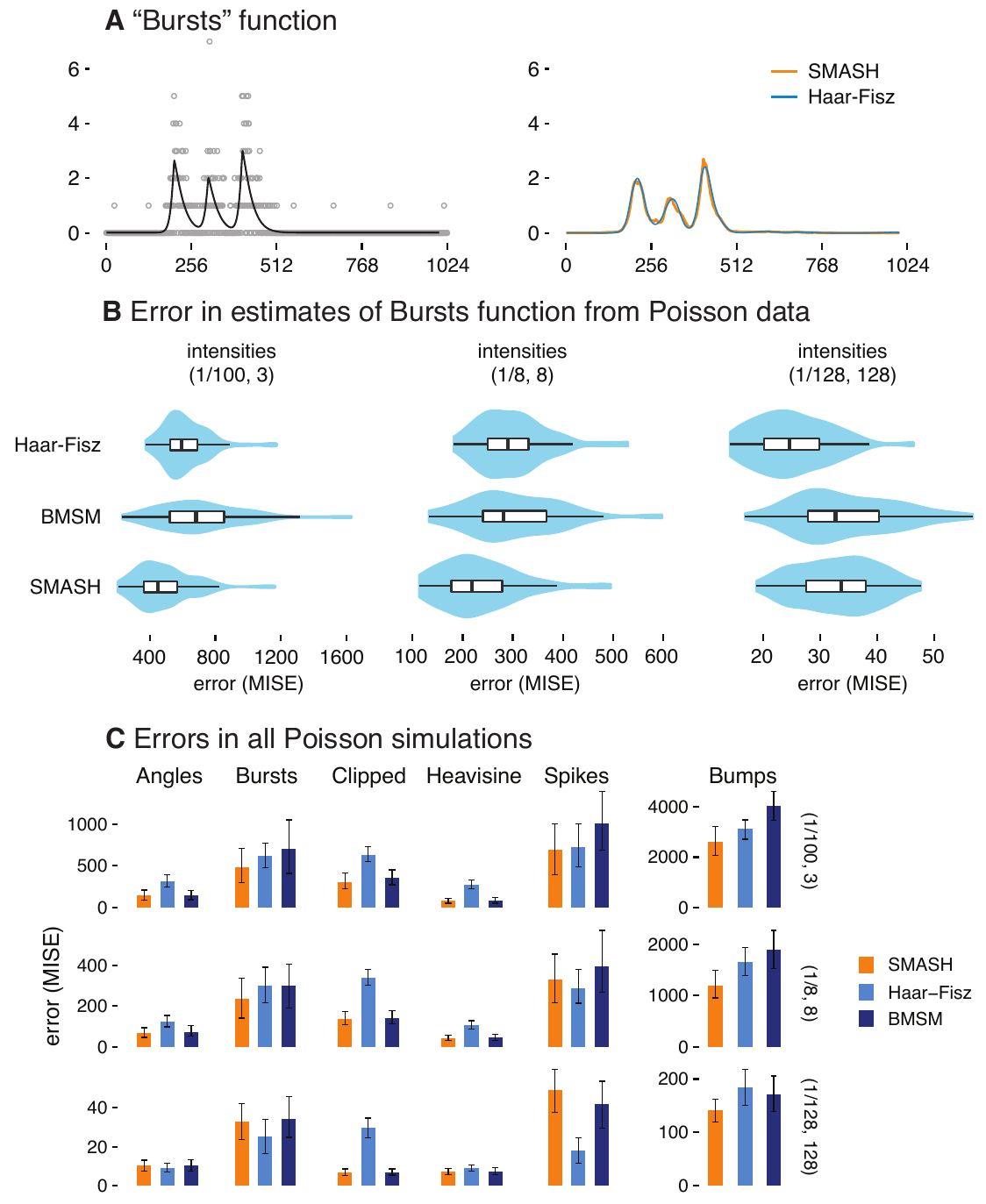}
\caption{Comparison of signal noising methods in Poisson data sets
  simulated with a variety of test functions and intensity ranges. For
  illustration, Panel A shows the ``Bursts'' test function (black
  line) and an example data set (gray circles) which was simulated at
  the $(1/100,3)$ range of intensities. The reconstructed signals
  (\mysmash{}, \textcolor{darkorange}{orange line}; HF,
  \textcolor{dodgerblue}{light blue line}) for this example data set
  are also shown. Panel C summarizes the error (MISE) in the mean
  estimates for all simulations; error bars show the 10\% and 90\%
  quantiles. The test functions used to simulate the data sets are
  shown in Figure \ref{fig:intensity_functions}. (Note the results for
  the ``Bumps'' simulations are plotted at a different scale because
  the MISE is much higher in these simulations.) For each of the
  scenarios, a total of 100 data sets were simulated at each intensity
  range, $(1/100,3)$, $(1/8,8)$ and $(1/128, 128)$. Methods compared
  are \mysmash, BMSM \citep{Kolaczyk1999Bayesian}, and the Haar-Fisz
  method \citep{Fryzlewicz2004HaarFisz} with a non-decimated wavelet
  transform and universal thresholding. Panel B gives a more detailed
  summary of the results from the ``Bursts'' simulations.}
\label{fig:pois_sim}
\end{figure}

In our final set of simulations, we assessed the ability of different
methods to reconstruct a spatially structured signal from
Poisson-distributed data. Similar to the Gaussian simulations, we
generated data sets using a variety of test functions and intensity
ranges. Specifically, we considered 6 test functions from
\cite{Besbeas2004Comparative, Fryzlewicz2004HaarFisz,
  Timmermann1999Multiscale} (see
Figure~\ref{fig:intensity_functions}), and defined $\bm{\mu}$ by
rescaling the test function so that the smallest intensity was $x$ and
the largest intensity was $y$, with $(x,y)$ set to either $(1/100,
3)$, $(1/8, 8)$ or $(1/128, 128)$. For each combination of test
function and intensity range, we simulated 100 data sets, each with a
signal of length $T = \mbox{1,024}$. We measured the accuracy of the
estimates using the mean integrated squared error (MISE), as we did above.

We compared \mysmash{} against the Bayesian multiscale model (BMSM)
and Haar-Fisz (HF) methods. BMSM is an Empirical Bayes method, like
\mysmash{}, but with a less flexible prior distribution on the
multi-scale coefficients \citep{Kolaczyk1999Bayesian}. The Haar-Fisz
method \citep{Fryzlewicz2004HaarFisz} performs a transformation of the
Poisson counts, then applies Gaussian wavelet methods to the
transformed data. There are many choices for Gaussian wavelet methods,
and the performance of the HF method is strongly dependent on which
Gaussian wavelet method is chosen, with different choices being better
for different data sets. We evaluated the performance of four variants
of the HF method---the details are given in Appendix
\ref{app:haar-fisz-variants}. Based on our empirical comparisons,
we found that the HF method with Gaussian denoising implemented using the
non-decimated wavelet transform and universal thresholding
\citep{Donoho1994Ideal}, and with a fixed noise level, yielded the
best estimates in most simulation scenarios, so in our results we
report results from the HF method with these settings.

The results of these simulations are summarized in
Figure~\ref{fig:pois_sim} (with additional figures and tables 
giving more detailed results
for all simulation settings included in the companion repository). 
In almost all simulation scenarios, \mysmash{}
performed as well or better than the HF and BMSM methods, 
with the greatest
gains occurring in the more challenging, lower intensity
scenarios. The only scenario where \mysmash{} was clearly outpeformed by another
method was the spikes scenario with high intensity range, where the HF method outpeformed both other methods. Comparing
BMSM with HF, neither dominated the
other: sometimes the BMSM method was better, whereas sometimes the HF method was better. As noted above, the HF transform can be used in a
variety of ways, so results here should be viewed only as a guide to
potential performance.

One practical limitation of the HF transform is that, to achieve
translation invariance, the transform has to be done explicitly for
each shift of the data: the tricks usually used to do this efficiently
\citep{Coifman1995Translationinvariant} do not work here. Thus, making
HF fully translation invariant increases computation by a factor of
$T$, rather than the factor of $\log(T)$ as for the other methods.  We
followed the advice of \citet{Fryzlewicz2004HaarFisz} and reduced the
computational burden by averaging over 50 shifts of the data rather
than $T$ shifts. With this approximation, the HF method was slower
than the other methods, but not by a lot. A direct comparison of
computational efficiency between \mysmash{} and BMSM is difficult as
they are coded in different programming environments. Nevertheless,
similarities between the two methods suggest that they should have
similar computational cost. In our simulations, the runtime of all
three methods was typically a few seconds or less per data set.

\subsection{Illustrative Applications}
\label{sec:applications}

In the experiments above, we showed that \mysmash{} is accurate for
denoising signals in simulated data sets, where the ground-truth
signal is known. To further illuminate the features of \mysmash{}, we
used \mysmash{} in two applications: analysis of motorcycle
acceleration data, which has been studied in other wavelet denoising
papers \citep{Delouille2004Smooth, silverman1985some}; and a problem
from computational biology---calling ``peaks'' in chromatin
immunoprecipitation sequencing (``ChIP-seq'') data \citep{chipseq,
  encode}.

\subsubsection{Motorcycle Acceleration Data}

\begin{figure}[t!]
\centering
\includegraphics[width=4.5in]{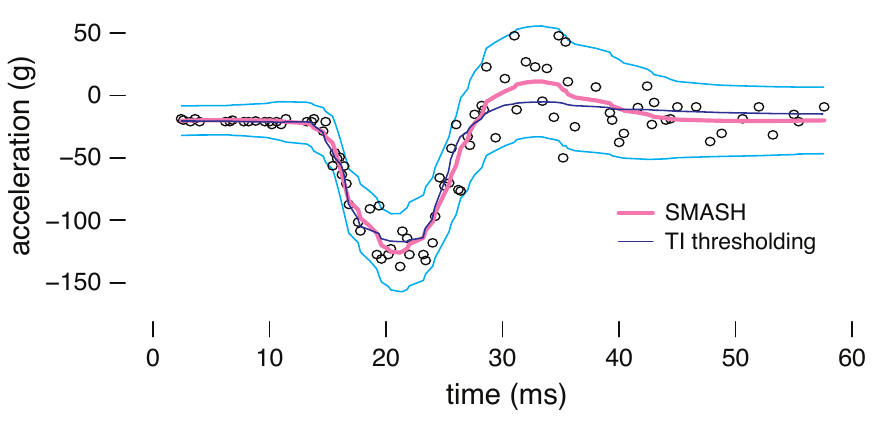}
\caption{\mysmash{} and TI thresholding applied to the motorcycle
  acceleration data \citep{silverman1985some}. The
  \textcolor{darkblue}{dark blue line} shows the signal recovered by
  TI thresholding, with RMAD estimates of the variance, and the
  \textcolor{pink}{pink line} shows the mean curve estimated by
  \mysmash. The $\pm2$ estimated standard deviations shown as the
  (\textcolor{dodgerblue}{dashed, light blue lines}). The data points
  are shown as {\bf black circles} ($\circ$).}
\label{fig:motorcycle}
\end{figure}

Here we demonstrate application of \mysmash{} to the motorcycle
acceleration data set from \citet{silverman1985some}. We chose this
data set because it exhibits clear heteroskedacity, and because it has
previously been found to be a challenging data set for wavelet
methods; for example, \citet{Delouille2004Smooth} required {\it ad
  hoc} data processing steps, including filtering out the
high-resolution wavelet coefficients, to produce an appealing fit.

The data consist of 133 observations measuring head acceleration from
a simulated motorcycle accident that was used to test crash
helmets. The dependent variable is acceleration (in {\it g}), and the
independent variable is time (in {\it ms}). To deal with repeated
measurements, we took the median of multiple acceleration measurements
at each time point. As in the analysis of Section~\ref{sec:mfvb}, we
treated the data as if they were equally spaced. In
  this example, we compare \mysmash{} to TI thresholding with RMAD
  variance estimates, since this method tended to be competitive with
  \mysmash{} in scenarios where changes to the variance were more
  gradual (Section~\ref{sec:het-gaussian-data}). This example is
  implemented by the ``Motorcycle Acceleration'' analysis in the
  online companion code repository, which includes a comparison with
  other variants of TI thresholding and \mysmash{} that are not shown
  here.

  The fitted \mysmash{} and TI thresholding curves are
  shown in Figure~\ref{fig:motorcycle}. Without hand-tuning of any
  parameters, both methods provide a reasonable fit to the
  data. Visually, \mysmash{} appears to favour a closer fit, whereas
  TI thresholding produces a slightly smoother curve. The
  nonparametric regression methods in \cite{Delouille2004Smooth} have
  more difficulty dealing with this data set (see Figure~11 of that
  paper).


\subsubsection{ChIP-seq Data}

Chromatin immunoprecipitation sequencing (``ChIP-seq'') is a widely
used technique to measure transcription factor binding along the
genome \citep{chipseq}. After preprocessing steps, the data are counts
of sequencing reads mapped to locations along the genome. These counts
can be treated as arising from an inhomogeneous Poisson process whose
intensity at site $b$ is related to the binding strength of the
transcription factor near $b$ \citep{anders2010, marioni2008}. Binding
tends to be localized---the vast majority of counts are expected to be
zero, with a small number of strong ``peaks''. Identifying these peaks
can help to identify regions where binding occurs, which is an
important component to understanding gene regulation. Consequently,
there are many methods for detecting ``peaks'' in ChIP-seq data
\citep{Wilbanks2010Evaluation}. Our goal here is to briefly describe
how \mysmash{} could provide an alternative approach to analyzing
ChIP-seq data by first estimating the underlying intensity function.
Once the intensity function has been estimated, ``peaks'' can be
identified as regions where the estimated intensity function exceeds
some predetermined threshold.

\begin{figure}[th!]
\centering
\includegraphics[width=5.4in]{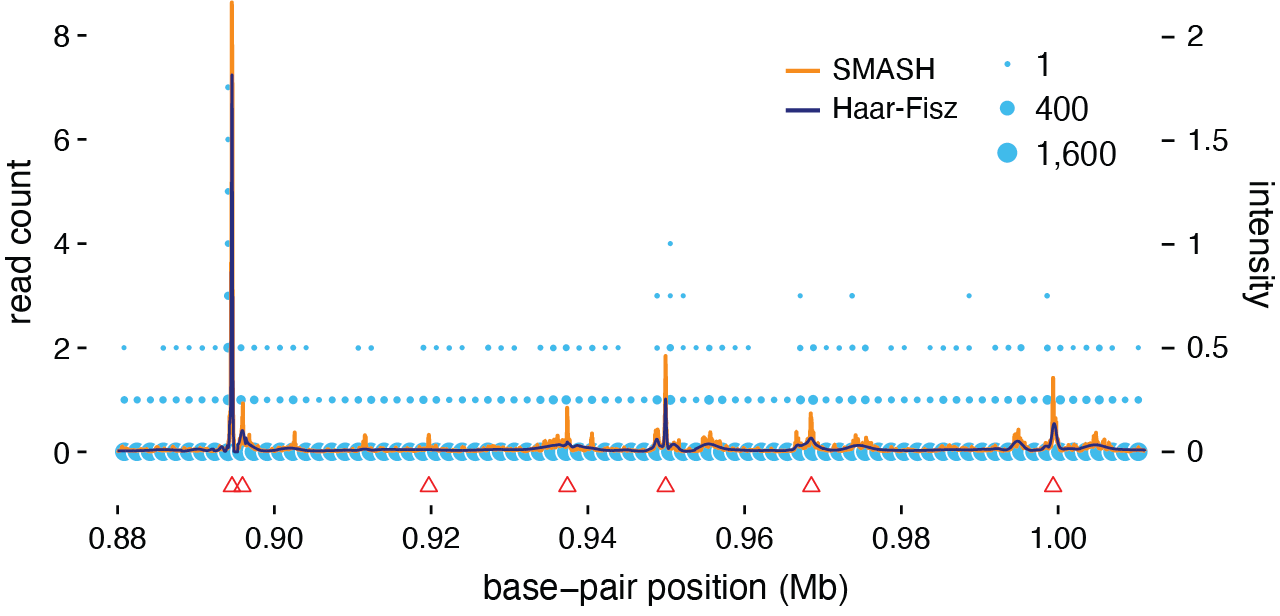}
\caption{Illustration of our \mysmash{}-based method for identifying
  peaks in ChIP-seq data. The data are ChIP-seq read counts for
  transcription factor {\it YY1} in cell line GM12878 from the ENCODE
  project \citep[``Encyclopedia of DNA
    Elements'';][]{encode-plos-biology, encode, encode-portal,
    gertz-2013, landt-2012}. Since this cell line has two ChIP-seq
  replicates (GEO accessions GSM803406 and GSM935482), the final
  counts were obtained by summing the read counts from both
  replicates. The region analysed comprises base-pair positions
  880,001--1,011,072 on chromosome 1, a region of $2^{17} \approx
  \mbox{131,000}$ base-pairs in length. (Base-pair positions are based
  on human genome reference assembly 19, NCBI build 37.)  Count data
  are depicted as \textcolor{dodgerblue}{light blue} circles, with
  circle area scaled by the number of data points within each 1.6-kb
  bin. (Note that most counts are zero.) The
  \textcolor{darkorange}{orange line} shows the intensity function,
  $\bm{\mu}$, estimated by \mysmash{}, and the
  \textcolor{darkblue}{dark blue line} shows the intensity function
  estimated by the HF method. MACS peaks \citep{Zhang2008Modelbased}
  are shown as \textcolor{red}{red triangles}
  (\textcolor{red}{$\bigtriangleup$}). (These are the mean positions
  of the MACS peak intervals.) This example is implemented by the
  ``Chipseq'' analysis in the accompanying source code repository.}
    \label{fig:seq_peak}
\end{figure}

To illustrate the approach, we applied \mysmash{} to a ChIP-seq data
set collected as part of the ENCODE project \citep[``Encyclopedia of
  DNA Elements'';][]{encode}. The data are ChIP-seq read counts at
$2^{17} \approx \mbox{131,000}$ locations (base-pair positions on
chromosome 1). The signal is very sparse; over 98\% of the read counts
(128,999 out of 131,072 base-pair positions) are zero. The \mysmash{}
analysis consists of estimating the mean and variance of the
underlying signal at these $2^{17}$ sites. For
  comparison, we also applied the Haar-Fisz method to these data
  (using the same settings used in Section~\ref{sec:poisson-data}).
  The \mysmash{} and HF methods each took about 5 minutes to run on
  these data (MacBook Pro, 3.5 GHz Intel i7 multicore CPU, R 3.4.3, no
  multithreaded external BLAS/LAPACK libraries).

  The intensity functions, $\bm{\mu}$, estimated by
  \mysmash{} and the HF method are shown in
  Figure~\ref{fig:seq_peak}. These estimates (the orange and dark blue
  lines) are overlaid with the ChIP-seq peaks (red triangles)
  identified by a widely used peak-calling software, MACS
  \citep{Zhang2008Modelbased}. The locations with the strongest
  intensity estimates align closely with the peaks found by
  MACS. However, the HF method recovered fewer MACS peaks, and at a
  much reduced intensity. The \mysmash{} estimates also suggest the
  presence of several additional weaker peaks not identified by MACS.

Reliable calling of peaks in ChIP-seq data is a multi-faceted problem,
and a full assessment of the potential for \mysmash{} to be applied to
this problem lies outside the scope of this paper.  Nonetheless, these
results suggest that this approach could be worth pursuing. One
benefit of our multi-scale Poisson approach is that it deals well with
a range of intensity functions, and could perform well even in
settings where peaks are broad or not well-defined. By contrast, the
performance of different peak-finding algorithms is often reported to
be sensitive to the ``kinds'' of peak that are present
\citep{Wilbanks2010Evaluation}. Therefore, developing peak-finding
algorithms that perform well in a range of settings remains an open
research question.

\section{Discussion}

We have introduced ``SMoothing by Adaptive SHrinkage'' (\mysmash{})
for smoothing Gaussian and Poisson data using multi-scale methods.
The method is built on the Empirical Bayes shrinkage method, \ash{},
whose two key features are: (i) models the multi-scale wavelet
coefficients using a flexible family of unimodal distributions; and
(ii) accounts for varying precision among coefficients. The first
feature allows \ash{} to flexibly adapt the amount of shrinkage to the
data, so data that ``look smooth'' are more strongly smoothed than
data that do not. The second feature allows \ash{} to deal effectively
with heteroskedastic variances, and consequently the mean gets
smoothed more strongly in regions where the variance is greater.

Notably, and unlike many wavelet shrinkage approaches, \mysmash{} is
self-tuning, and requires no specification of a ``primary resolution
level'' \citep[e.g.,][]{Nason2002Choice} or other tuning
parameters. This feature is due to the ``adaptive'' nature of \ash{}
noted above; when a particular resolution level shows no strong signal
in the data, \ash{} learns this and adapts the amount of shrinkage
(smoothing) appropriately. This ability to self-tune is important for
two reasons. First, it makes the method easier to use by non-experts,
who may find appropriate specification of tuning parameters
challenging. Second, it means that the method can be safely applied
``in production'' to large numbers of data sets in settings such as
genomics where it is impractical to hand-select appropriate tuning
parameters separately for every data set.

Our results here demonstrate that \mysmash{} provides a flexible, fast
and accurate approach to smoothing and denoising. We illustrated this
flexibility by applying it to two challenging problems: Gaussian
heteroskedastic regression and smoothing of Poisson signals. In both
cases, our method is competitive with existing approaches.

While \mysmash{} requires more computation than a simple thresholding
rule, it is fast enough to deal with large problems. This is partly
because fitting the unimodal distribution in \ash{} is a convex
optimization problem that can be solved stably and quickly using
existing numerical optimization techniques \citep{mixsqp,
  koenker2015rebayes, stephens2017false}. Using the convex
optimization library MOSEK \citep{mosek}, which is interfaced through
the ``KWDual'' function in the R package {\tt REBayes}
\citep{koenker2015rebayes}, fitting the \ash{} model typically takes
about 30 seconds or less for a data set with 100,000
observations. (This timing is based on running R 3.4.3 on a MacBook
Pro with a 3.5 GHz Intel i7 multicore CPU and no multithreaded
external BLAS/LAPACK libraries.) \mysmash{} requires multiple
applications of \ash{}---it is applied at each resolution level, and
requires $\log_2(T)$ applications in the Poisson case---yet it remains
fast enough to be practical for moderately large problems; for
example, smoothing a signal of length $2^{15}$ = 32,768 typically
takes less than 1 minute for Poisson-distributed data, and less than 2
minutes for Gaussian data. It is likely these runtimes could be
further improved by more efficient implementations.

Besides its accuracy for point estimation, \mysmash{} also has the
advantage that it naturally provides measures of uncertainty in
estimated wavelet coefficients, which in turn provide measures of
uncertainty (e.g., credible bands) for estimated mean and variance
functions.

Although we have focussed here on applications in one dimension,
\ash{} could potentially be applied to multi-scale approaches in
higher dimensions, such as image denoising
\cite{Nowak1999Multiscale}. Alternatives to wavelets, such as
curvelets \citep{Cande00}, may produce better results for image
processing applications. Extending our work to those settings could be
an interesting direction for future work.

\section*{Acknowledgements}

This work was supported in part by NIH grant HG002585 to MS. We thank
the ENCODE Consortium, R. Myers and F. Pauli at HudsonAlpha, and
P. Cayting at the Stanford Center for Genomics and Personalized
Medicine making the ChIP-seq data available.

\def\ff{f}

\appendix

\section{Variance Estimation for Gaussian Denoising}
\label{app:var_estimation}

With $\bm{Z}$ as defined in \eqref{eqn:varobs1}, we apply the wavelet
transform $W$ to $\bm{Z}^2$, and obtain the wavelet coefficients
$\bm{\Gd} = W\bm{Z}^2$. Note that $\EE(\bm{\Gd}) = \bm{\Gg}$, where
$\bm{\Gg} = W\bm{\s}^2$. We treat the likelihood for $\bm{\Gg}$ as if
it were independent, resulting in
\begin{equation*}
L(\bm{\Gg}\,|\,\bm{\Gd}) =
  \prod_{j=0}^J\prod_{k=0}^{T-1} p(\Gd_{jk}\,|\,\Gg_{jk}).
\end{equation*}
The likelihoods $L(\Gg_{jk} \,|\,\Gd_{jk})$ are not normal, but we
approximate the likelihood by a normal density through matching the
moments of a normal distribution to the distribution
$p(\Gd_{jk}\,|\,\Gg_{jk})$; that is,
\begin{equation*}
p(\Gd_{jk}\,|\,\Gg_{jk}) \approx N(\Gg_{jk},\hat{s}^2(\Gd_{jk}))
\end{equation*}
so that
\begin{eqnarray*}
L(\Gg_{jk}\,|\,\Gd_{jk}) \approx
  \phi(\Gd_{jk};\Gg_{jk},\hat{s}^2(\Gd_{jk})),
\end{eqnarray*}
where $\phi$ is the normal density function, and $\hat{s}^2(\Gd_{jk})$
is the variance of the empirical wavelet coefficients. Since these
variances are unknown, we estimate them from the data and then proceed
to treat them as known. Specifically, since $Z_t\sim N(0,\s_t^2)$, we
have that
\begin{eqnarray*}
&\EE(Z_t^4)\approx 3\s_t^4 \\
& \mathrm{Var}(Z_t^2)\approx 2\s_t^4,
\end{eqnarray*}
so we simply use $\frac{2}{3}Z_t^4$ as an unbiased estimator for
$\mathrm{Var}(Z_t^2)$. It then follows that $\hat{s}^2(\Gd_{jk})$ is
given by $\sum_{l=1}^T \frac{2}{3}Z_l^4W_{jk,l}^2$, and is an unbiased
estimte of $\mathrm{Var}(\Gd_{jk})$. These will be the inputs to
\ash{}, which then produces shrunk estimates in the form of posterior
means for the corresponding parameters. Although this works well in
most cases, there are variance functions for which the above procedure
tends to overshrink the wavelet coefficients at the finer levels. This
is likely because the distribution of the wavelet coefficients is
extremely skewed, especially when the true coefficients are small (at
coarser levels the distributions are much less skewed since we are
dealing a linear combination of a large number of data points). One
way around this issue is to employ a procedure that jointly shrinks
the coefficients $\bm{\Gg}$ and their variance estimates (this is
implemented by the \verb|jash| option in our software). The final
estimate of the variance function is obtained from the posterior means
via the average basis inverse across all the shifts.

\section{Poisson Denoising}
\label{app:reconstruction}

First, we summarize the data in a recursive manner by defining
\begin{eqnarray*}
y_{J,k}\equiv y_k,
\end{eqnarray*}
for $k = 1,\ldots,T$, with $T=2^J$, and
\begin{eqnarray*}
y_{jk} = y_{j+1,2k} + y_{j+1,2k+1}
\end{eqnarray*}
for resolutions $j=0,\ldots,J-1$ and locations $k=0,\ldots,2^j-1$. Hence, we
are summing more blocks of observations as we move to coarser levels.

This recursive scheme leads to:
\begin{eqnarray*}
y_{jk} = \sum_{l = k2^{J-j}+1}^{(k+1)2^{J-j}} y_l
\end{eqnarray*}
for $j=0,\ldots,J$ and $k=0,\ldots,2^j-1$.

Similarly, we define
\begin{eqnarray*}
\mu_{J,k}\equiv \mu_k
\end{eqnarray*}
for $k=1,\ldots,T$, and
\begin{eqnarray*}
\mu_{jk} = \mu_{j+1,2k} + \mu_{j+1,2k+1}
\end{eqnarray*}
for $j=0,\ldots,J-1$ and $k=0,\ldots,2^j-1$. 
And define
\begin{eqnarray*}
\Ga_{jk}=\log \mu_{j+1,2k} - \log \mu_{j+1,2k+1}
\end{eqnarray*}
for $s=0,\ldots,J-1$ and $l=0,\ldots,2^j-1$. The $\Ga_{jk}$'s defined
this way are analogous to the (true) Haar wavelet coefficients for
Gaussian signals.

Using this recursive representation, the likelihood for $\bm{\Ga}$
factorizes into a product of likelihoods, where $\bm{\Ga}$ is the
vector of all the $\Ga_{jk}$'s. See \citet{Kolaczyk1999Bayesian}, for
example. Specifically,
\begin{align*}
L(\bm{\Ga} \,|\, \mathbf{Y}) &= p(\mathbf{Y} \,|\, \bm{\Ga})\\
&= p(y_{0,0} \,|\, \mu_{0,0}) \,
   \prod_{j=0}^{J-1}\prod_{k=0}^{2^j-1} p(y_{j+1,2k} \,|\, y_{j,k},\Ga_{j,k})\\
&= L(\mu_{0,0} \,|\, y_{0,0}) \,
   \prod_{j=0}^{J-1}\prod_{k=0}^{2^j-1}L(\Ga_{j,k} \,|\, y_{j+1,2k},y_{j,k}).
\end{align*}
Note that $y_{00} \,|\, \mu_{00} \sim \Poi(\mu_{00})$. For any
given $j$ and $k$, $y_{jk}$ is a sum of two independent Poisson random
variables, and is itself a Poisson random variable. Hence,
\begin{equation*}
y_{j+1,2k} \,|\, y_{jk}, \Ga_{jk} \sim
\Bin\Big(y_{jk},\frac{1}{1+e^{-\Ga_{jk}}}\Big) =
\Bin\Big(y_{jk},\frac{\mu_{j+1,2k}}{\mu_{jk}}\Big).
\end{equation*}

\subsection{Estimates and Standard Errors for $\alpha_j$}

Each $\alpha_{j}$ is a ratio of the form $\log(\mu_{a:b}/\mu_{c:d})$,
whose maximum likelihood estimate (MLE) is $\log(y_{a:b}/y_{c:d})$.
The main challenge here is that the MLE is not well behaved when
either the numerator $y_{a:b}$ or denominator $y_{c:d}$ is zero. To
deal with the case when either is zero, we use Tukey's modification
\citep{Gart1967Bias}. Specifically, letting $S$ denote $y_{a:b}$, $F$
denote $y_{c:d}$ and $N = S+F$ (effectively treating these as
successes and failures in a binomial experiment, conditioned on
$y_{a:b} + y_{c:d}$), we use estimator
\begin{align}
\label{eqn:pseudoMLE1}
\hat{\Ga}&=\left\{
\begin{array}{lll}
\log\{(S+{\textstyle\frac{1}{2}})/(F+{\textstyle\frac{1}{2}})\} -
      {\textstyle\frac{1}{2}}&  \mbox{if $S=0$} \\[0.75ex]
\log\{(S + {\textstyle\frac{1}{2}})/(F + {\textstyle\frac{1}{2}})\} +
      {\textstyle\frac{1}{2}} & \mbox{if $S=N$}\\[0.75ex]
\log(S/F) & \mbox{otherwise} \\
\end{array}
\right.\\
\label{eqn:pseudoMLE1se}
se(\hat{\Ga}) &= \sqrt{V^*(\hat{\Ga}) -
  {\textstyle\frac{1}{2}}V_3(\hat{\Ga})^2
  \left(V_3(\hat{\Ga})-{\textstyle\frac{4}{N}}\right)},
\end{align}
where
\begin{align*}
V_3(\hat{\Ga}) &= {\textstyle \frac{N+1}{N}}
\left(\textstyle{\frac{1}{S+1}} + {\textstyle\frac{1}{F+1}}\right),
\qquad (S = 0,\ldots,N) \\
V^*(\hat{\Ga}) &=
V_3(\hat{\Ga})\left(1 - {\textstyle\frac{2}{N}} +
                   {\textstyle\frac{V_3(\hat{\Ga})}{2}}\right).
\end{align*}
The square of the standard error in \eqref{eqn:pseudoMLE1se}
corresponds to $V^{\ast\ast}$ from p.~182 of \citet{Gart1967Bias}, and
is chosen because it is less biased for the true variance of
$\hat{\Ga}$ (when $N$ is small) as compared to the asymptotic variance
of the MLE \citep[see][]{Gart1967Bias}. The other two variance
estimators from \citet{Gart1967Bias}, $V_1^{++}$ and $V^{++}$, were
also considered in simulations and gave similar results, but
$V^{\ast\ast}$ was chosen for its simpler form.

\subsection{Signal Reconstruction}

The first step to reconstructing the signal is to find the posterior
means of $p_{jk} \colonequals \frac{\mu_{j+1,2k}}{\mu_{jk}}$ and $q_{jk}
\colonequals \frac{\mu_{j+1,2k+1}}{\mu_{jk}}$, for $j=0,\ldots,J-1$ and
$k=0,\ldots,2^j-1$. Specifically, for each $j$ and $k$, we require
\begin{align}
  \label{eqn:pfromwc1}
\EE(p_{jk}) &\equiv \EE\left(\frac{1}{1+e^{-\Ga_{jk}}}\right) \\
\label{eqn:pfromwc2}
\EE(q_{jk}) &\equiv \EE\left(\frac{1}{1+e^{\Ga_{jk}}}\right).
\end{align}
Given the posterior means and variances for $\Ga_{jk}$ from \ash{}, we
can approximate (\ref{eqn:pfromwc1}--\ref{eqn:pfromwc2}) using the
delta method. First, we define
\begin{equation*}
\ff(x) = \frac{1}{1-e^{-x}},
\end{equation*}
and consider the Taylor expansion of $\ff(x)$ about $\ff(\EE(x))$,
\begin{eqnarray*}
\ff(x)\approx \ff(\EE(x))+ d\ff(\EE(x))(x - \EE(x)) +
\frac{d^2\!f(\EE(x))}{2}(x - \EE(x))^2,
\end{eqnarray*}
where
\begin{align*}
d\ff(x) &= \frac{e^x}{(1+e^x)^2}\\
d^2\!\ff(x) &= \frac{e^x(1-e^{x})}{(1+e^x)^3}.
\end{align*}
Therefore,
\begin{align*}
E(p_{jk}) &\approx
  \ff(E(\Ga_{jk}))+\frac{d^2\!\ff(E(\Ga_{jk}))}{2}\mathrm{Var}(\Ga_{jk}) \\
E(q_{jk}) &\approx
  \ff(-E(\Ga_{jk}))+\frac{d^2\!\ff(-E(\Ga_{jk}))}{2}\mathrm{Var}(\Ga_{jk}),
\end{align*}
which can be computed by plugging in $\EE(\Ga_{jk})$ and
$\mathrm{Var}(\Ga_{jk})$ from \ash{}.

Finally, we approximate the posterior mean for $\mu_t$ by noting that
$\mu_t$ can be written as a product of the $p_{jk}$'s and $q_{jk}$'s
for any $t=1,2,\ldots,T$. Specifically, let $c_1, \ldots, c_J$ be the
digits of the binary encoding of $t-1$, and let $d_m = \sum_{j=1}^m
c_j2^{m-j}$, for $j=1,\ldots,J-1$. Then we have that
\begin{eqnarray}
\label{eqn:product}
\mu_t = \mu_{00} \, p_{00}^{1-c_1} \, p_{1,d_1}^{1-c_2} \cdots
  p_{J-1,d_{J-1}}^{1-c_J} \, q_{00}^{c_1} \, q_{1,d_1}^{c_2} \cdots
  q_{J-1,d_{J-1}}^{c_J},
\end{eqnarray}
where we usually estimate $\mu_{00}$ as $\sum_l y_l$, following
\citet{Kolaczyk1999Bayesian}. Further, exploiting the independence of the
$p_{jk}$'s and $q_{jk}$'s at different scales, we have that
\begin{align}
\label{eqn:Eproduct}
\EE(\mu_t) =& \mu_{00} \, \EE(p_{00})^{1-c_1} \EE(p_{1,d_1})^{1-c_2} \cdots
\EE(p_{J-1,d_{J-1}})^{1-c_J} \nonumber \\
& \qquad \times \EE(q_{00})^{c_1} \, E(q_{1,d_1})^{c_2} \cdots
\EE(q_{J-1,d_{J-1}})^{c_J}.\end{align}

We can also approximate the posterior variance of $\mu_t$. (This
allows creation of an approximate credible interval under normal
approximation.)  From \eqref{eqn:product}, we have
\begin{align}
  \label{eqn:E2product}
\EE(\mu_t^2) =& \mu_{00}^2E(p_{00}^2)^{1-c_1} \, \EE(p_{1,d_1}^2)^{1-c_2}
  \cdots \EE(p_{J-1,d_{J-1}}^2)^{1-c_J} \nonumber \\
& \qquad \times \EE(q_{00}^2)^{c_1} \, \EE(q_{1,d_1}^2)^{c_2} \cdots
  \EE(q_{J-1,d_{J-1}}^2)^{c_J}.
\end{align}
To compute this quantity, we again use the delta method, with $\ff(x)
= \big(\frac{1}{1+e^{-x}}\big)^2$, to obtain:
\begin{align}
\EE(p_{jk}^2) &\approx \big(\ff(E(\Ga_{jk})) +
d^2\!\ff(\EE(\Ga_{jk})) \mathrm{Var}(\Ga_{jk})/2\big)^2 +
\{d\ff(\EE(\Ga_{jk}))\}^2 \mathrm{Var}(\Ga_{jk}) \\
\EE(q_{jk}^2) &\approx \big(\ff(-\EE(\Ga_{jk})) +
d^2\!\ff(-E(\Ga_{jk})) \mathrm{Var}(\Ga_{jk})/2\big)^2 
+ \{d\ff(\EE(-\Ga_{jk}))\}^2 \mathrm{Var}(\Ga_{jk}).
\end{align}
Finally, we combine \eqref{eqn:Eproduct} and \eqref{eqn:E2product} to
obtain $\mathrm{Var}(\mu_t)$.

\subsection{Translation Invariance}
\label{app:TI}

It is common in multi-scale analysis to perform analyses over all $T$
circulant shifts of the data, because this is known to consistently
improve accuracy. (The $t$-th circulant shift of the signal $\bm{Y}$
is created from $\bm{Y}$ by moving the first $T-t$ elements of
$\bm{Y}$ $t$ positions to the right, then inserting the last $t$
elements of $\bm{Y}$ into the first $t$ locations.)

To implement this in practice, we begin by computing the $\Ga_j$
coefficients, and their corresponding standard errors, for all $T$
circulant shifts of the data. This is done efficiently in $O(\log_2
T)$ operations using ideas from
\citet{Coifman1995Translationinvariant}. We took the steps described in
\citet{Kolaczyk1999Bayesian}; indeed, our software implementation
benefitted from MATLAB code provided by \citet{Kolaczyk1999Bayesian}
for the TI table construction, which we ported to C++ and interfaced
to R using {\tt Rcpp} \citep{eddelbuettel2011rcpp}.

This yields a table of $\Ga$ coefficients, with $T$ coefficients at
each of $\log_2 T$ resolution levels, and a corresponding table of
standard errors. As in the Gaussian case, we then apply \ash{}
separately to the $T$ coefficients at each resolution level to obtain
a posterior mean and posterior variance for each $\Ga_j$. Finally, we
use the methods detailed above to compute quantities of interest
averaged over all $T$ shifts of the data. For example, our final
estimate of the mean signal $\mu_k$, for $k = 1, \ldots, T$, is given
by $\sum_{t=1}^T \hat{\mu}_k^{(t)}/T$, where $\hat{\mu}_k^{(t)}$
denotes the posterior mean of $\mu_k$ computed from the $t$-th
circulant shift of the data.  Again, borrowing ideas from
\citet{Coifman1995Translationinvariant}, this averaging can be done
with $O(\log_2 T)$ operations.

\section{Implementation of Haar-Fisz method in Poisson simulations}
\label{app:haar-fisz-variants}

We explored four options for the Gaussian denoising stage of the
Haar-Fisz method, all with 50 ``external cycle-spins''
\citep{Fryzlewicz2004HaarFisz}:

\begin{enumerate}
  
\item A hybrid of the greedy tree denoising algorithm
  \citep{Baraniuk1999Optimal} and wavelet thresholding using
  ``leave-half-out'' cross-validation \citep{Nason1995Choice}. We used
  $j_0 = 3$ (the default setting), and the noise level was estimated
  from the data. These choices correspond to the ``H:CV+BT CS'' method in
  \citep{Fryzlewicz2004HaarFisz}. In practice, we found that the
  algorithm did not always converge, in which case we marked the
  solution as being unavailable.

\item Wavelet thresholding using the universal threshold
  \citep{Donoho1994Ideal}. We used $j_0=3$ (the default setting),
  and the noise level was estimated from the data. These choices
  correspond to the ``F$\bowtie$U CS'' method in
  \citep{Fryzlewicz2004HaarFisz}.
  
\item Wavelet thresholding using the universal threshold for the
  non-decimated wavelet transform. Results were averaged over settings
  $j_0 = 4, 5, 6, 7$, and the noise level was estimated from the data.
  
\item Wavelet thresholding using the universal threshold for the
  non-decimated wavelet transform, in which the noise level was set to
  1 rather than estimating it from the data (this is the asymptotic
  variance under the Fisz transform). Results were averaged over
  settings $j_0 = 4, 5, 6, 7$.
  
\end{enumerate}
The settings of each HF method were chosen by us to optimize (average)
performance through moderately extensive experimentation on a range of
simulations.

\section{Test functions used to simulate data}

\begin{figure}[t!]
\centering
\includegraphics[width=5.5in]{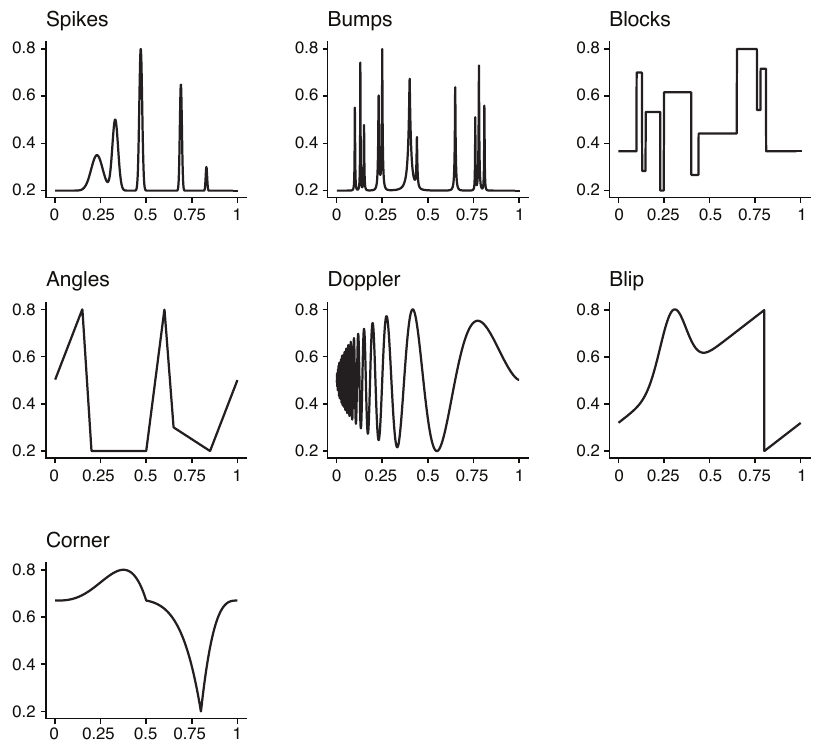}
\caption{Mean functions used to simulate the Gaussian data
  sets.} \label{fig:gaussian_mean_signals}
\end{figure}

\begin{figure}[t!]
\centering
\includegraphics[width=5.5in]{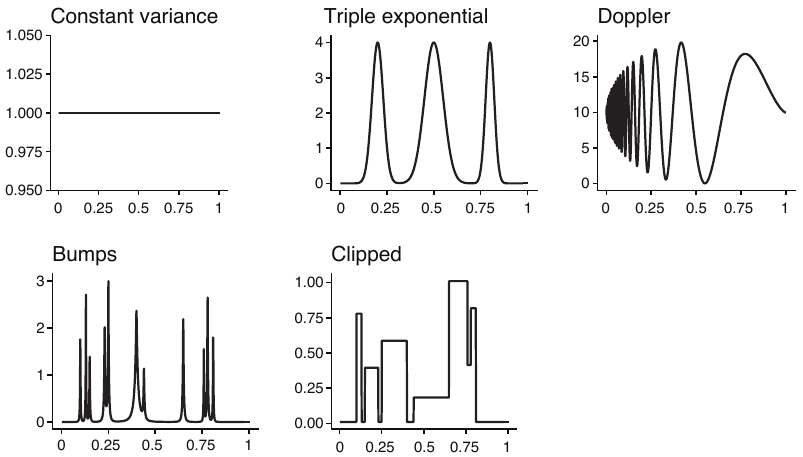}
\caption{Variance functions used to simulate the
    Gaussian data sets. In practice, these functions are rescaled in
    the simulations to achieve the desired signal-to-noise ratios.}
  \label{fig:gaussian_variance_signals}
\end{figure}

\begin{figure}[t!]
\centering
\includegraphics[width=5.5in]{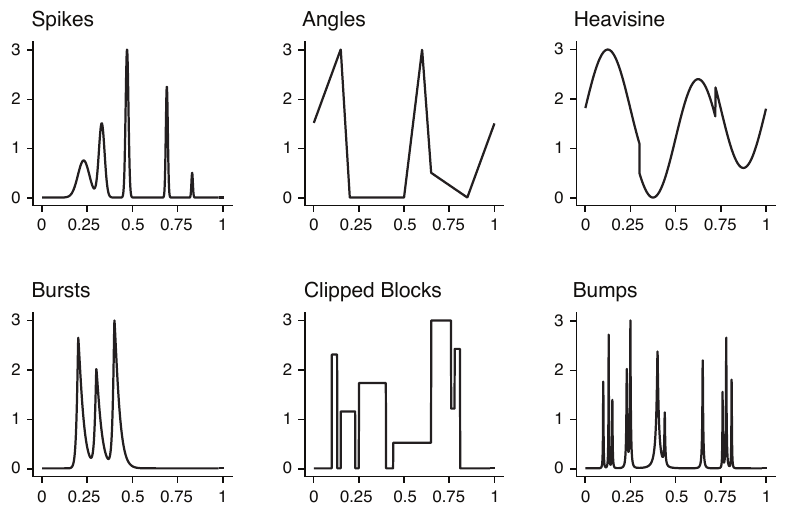}
\caption{Intensity functions used to simulate the Poisson 
data sets.}
\label{fig:intensity_functions}
\end{figure}

Figures \ref{fig:gaussian_mean_signals} and
\ref{fig:gaussian_variance_signals} show the mean and variance
functions used to simulate the Gaussian data sets. Figure
\ref{fig:intensity_functions} shows the intensity functions used to
simulate the Poisson data sets.

\bibliography{smash}

\end{document}